\documentclass[a4paper,11pt]{article}
\pdfoutput=1 

\usepackage{jheppub} 
\usepackage[dvipsnames]{xcolor}
\usepackage{amsmath}
\usepackage{amssymb}
\usepackage{bbm}
\usepackage{mathrsfs}  
\usepackage{bm}
\usepackage{multirow}
\usepackage[T1]{fontenc} 
\usepackage[normalem]{ulem}

\newcommand{\FSL}[1]{F_{#1}}
\newcommand{\FSLub}[1]{A_{#1}}
\newcommand{\FSLbr}[1]{R_{#1}}
\newcommand{\FSNLub}[1]{a_{#1}}
\newcommand{\FSNLbr}[1]{r_{#1}}
\newcommand{\FSMNLz}[1]{z_{#1}}
\newcommand{\FSMNLw}[1]{w_{#1}}
\newcommand{\htr}{\mathfrak{h}}
\newcommand{\mH}{\mathcal{H}}
\newcommand{\Rep}{\mathcal{R}}
\newcommand{\bpi}{\bm{\pi}}
\newcommand{\ngbmc}[1]{v_{#1}}
\newcommand{\ngbmcL}[1]{V_{#1}}
\newcommand{\OPM}{\mathscr{R}}
\newcommand{\dmu}{[d\bar\mu(q\partial)]}
\newcommand{\PE}{\textrm{PE}}

\newcommand{\Qbar}{\bar{Q}}
\newcommand{\Lbar}{\bar{L}}

\title{Counting and building operators in theories with hidden symmetries and application to HEFT}

\author{Rodrigo Alonso}
\author{and Shakeel Ur Rahaman}

\affiliation{Institute for Particle Physics Phenomenology, Durham University, Durham DH1 3LE, UK}

\emailAdd{rodrigo.alonso-de-pablo@durham.ac.uk}
\emailAdd{shakeel.u.rahaman@durham.ac.uk}
\preprint{IPPP/24/80}

\abstract{ Identifying a full basis of operators to a given order is key to the generality of Effective Field Theory (EFT) and is by now a problem of known solution in terms of the Hilbert series.
The present work is concerned with hidden symmetry in general and Higgs EFT in particular and {\it(i)} connects the counting formula presented in~\cite{Graf:2022rco} in the CCWZ formulation with the linear frame and makes this connection explicit in HEFT {\it (ii)} outlines the differences in perturbation theory in each frame {\it (iii)} presents a new counting formula with measure in the full $SU(3)\times SU(2)\times U(1)$ group for HEFT and {\it (iv)} provides a Mathematica code that produces the number of operators at the user-specified order in HEFT.}

\begin{document} 
\maketitle
\flushbottom

\section{Introduction}
\label{sec:intro}

The Standard Model (SM) has been extensively tested at the Large Hadron Collider (LHC) and has consistently aligned with experimental results. As a consequence of observations the theoretical expectation for new physics at the electroweak scale is under critical revision; while solid evidence for Beyond the Standard Model remains (e.g. Dark Matter (DM), neutrino masses or the Baryon Asymmetry of the universe), this might not manifest itself at the electroweak (EW) scale in the form predicted decades ago. In this context the need for a general encompassing framework arose and directed part of the theory community towards Effective Field Theory \cite{Weinberg:1978kz, Weinberg:1980wa, Georgi:1994qn, Manohar:2018aog, Manohar:1996cq}.

Provided the mass scale of BSM physics is beyond the direct reach of the LHC, one can use the same theory tool that describes nuclear beta decay mediated by the $W$ boson at nuclear scales. In place of Fermi's four fermion operator there will be many others in our EW scale study if one is to be comprehensive and comprehensiveness is what directed us here. This consideration of all possible terms is key since it grants independence of the specific microscopic theory, the $W$ boson in nuclear physics or in our case the ultra-violet scale out of reach at the LHC. To be precise, the generality comprises all UV models which are mapped via matching \cite{Gaillard:1985uh, Chan1985, Cheyette:1987qz, Henning:2014wua, Drozd:2015rsp, Ellis:2016enq, delAguila:2016zcb, Ellis:2017jns, Kramer:2019fwz, 
Banerjee:2023iiv, Banerjee:2023xak, Chakrabortty:2023yke, Cohen:2020fcu, Dittmaier:2021fls} and can be selected by a specific choice of coefficients in our operators.

What makes such a comprehensive program viable is an expansion parameter to organize our operators, together with the fundamental symmetries and field content. The realisation of the internal symmetry itself is not a straightforward matter; if $SU(3)_c\times SU(2)_w\times U(1)_Y$ is realised linearly and with a complete linear representation breaking the symmetry, as is the case in the SM, one obtains the Standard Model Effective Field Theory (SMEFT) \cite{Brivio:2017vri, Isidori:2023pyp}. This theory has a clear expansion parameter in a scale $\Lambda$ and leading lepton-number-preserving effects appear at order $\Lambda^{-2}$, a.k.a. dimension six operators which have been extensively studied \cite{Dawson:2021ofa, Dawson:2020oco, Ellis:2021jhep, Banerjee:2019twi, deBlas:2019qco, deBlas:2019rxi, Dawson:2020prd, Ellis:2018jhep, Biekotter:2023xle, Biekotter:2023vbh, Bartocci:2023nvp, Anisha:2021hgc}. The above formulation highlights the key assumption of SMEFT, the responsible field for the breaking is a linear doublet; this need not be the case indeed. Physical massive bosonic states are the spin one $W$, $Z$ and a scalar $h$ where the latter need not be related to the longitudinal states in the former as a doublet demands; dropping this assumption one reaches instead the more comprehensive Higgs EFT (HEFT) \cite{Alonso:2023upf, Feruglio:1992wf, Alonso:2012px, Alonso:2016oah, Helset:2020yio, Buchalla:2013rka}. The separation of Goldstones and the scalar $h$ while respecting the symmetry is achieved by the CCWZ \cite{Callan:1969sn, Coleman:1969sm} construction where only $SU(3)_c\times U(1)_{\rm em}$, i.e. the unbroken symmetry, is realised linearly while the broken part is encoded in GB-dependent nonlinear transformations. This CCWZ frame with non-linear symmetry can be related to the linear frame (in which one writes the SMEFT) since the Goldstone themselves are the interpolating field. It is the case in fact that even though the HEFT construction uses the essence of CCWZ, it is usually given in the linear frame.

The next step once the symmetry realisation is specified is to list all possible operators to a given order. This problem, despite being riddled by redundancies arising from equations of motion (EOMs), integration by parts (IBPs), Fierz identities and other relations,  has been solved with full generality both for SMEFT and HEFT.  The method used in this solution has come to be known as the  Hilbert Series (HS) and has been extensively used in the construction of different EFTs \cite{Lehman:2015via, Lehman:2015coa, Henning:2015alf, Henning:2017fpj, Graf:2020yxt, Graf:2022rco}.

This work aims at following up on the results of \cite{Graf:2022rco} by connecting explicitly the CCWZ and the linear frame used in the wider HEFT literature. We outline perturbation theory in each frame, presenting a new counting formula purely in the linear frame for HEFT and provide a Mathematica code that produces the number of operators at a desired order. For this purpose this paper presents first a pedagogical review of HS in sec.~\ref{ReviewHS} with the non-linear realisation in~\ref{RevNLHS} and the linear-CCWZ connection in~\ref{Connek}. The application to HEFT in the CCWZ frame is in~\ref{sec:frameCCWZ} and linear frame in~\ref{LinBasFor} together with the new counting formula. The treatment of Higgs-related redundancies and the three different counting formulas that follow are presented in sec.~\ref{sec:modding} while the code is presented in sec.~\ref{sec:code}.

\section{Hilbert series in QFT}\label{ReviewHS}
In this section we aim to provide an accessible and intuitive introduction to the use of Hilbert series in operator counting which will also serve to outline results that this work is based on. The exposition might differ from other works but not the content itself which can be found in e.g. \cite{Henning:2017fpj} with more rigour and comprehensiveness. 

At the centre of this method lies the character orthogonality theorem. Consider a group $G$, a group element $g$, and irreps $\Rep$, $\Rep'$; one has that the group average over the product of characters $\chi$ is orthogonal, i.e.
\begin{align}\label{OrTH}
\frac{1}{V_G}\sum_{g}\chi^*_{\Rep'}\chi_{\Rep}&=\delta_{\Rep\Rep'}\;,&\chi_\Rep&\equiv\mbox{Tr}(g_\Rep)\;,
\end{align}
where $V_G$ is the volume of the group (number of elements if discreet, integral of the Haar measure if continuous) and the subindex $\Rep$ in $g_\Rep$ denotes the group element as a matrix in $\Rep$ space. For any two representations one has that (a) $\chi_{\Rep_1\times \Rep_2}=\Sigma_{\rm irreps}\chi_{\Rep_i}$ while if the two representations correspond to distinct fields, one has further (b) $\chi_{\Rep_1\times \Rep_2}=\chi_{\Rep_1}\chi_{\Rep_2}$ as one can check explicitly. Given (a) one can use the orthogonality theorem in eq.~(\ref{OrTH}) to obtain the decomposition in irreducible representations of the combination $\Rep_1\otimes \Rep_2$ while (b) gives an explicit form for the average to be evaluated.
As an example, consider $N$ different fields each in a representation $\Rep_i$; the answer to the question, how many invariants can be built out of their product is $n_I$
\begin{align}
\Rep_1&\otimes \Rep_2\otimes\dots \otimes \Rep_N=(n_I\times \mathbf{1})\oplus \dots\;,\\
  n_I& = \frac{1}{V_G}\sum_g \chi_{\mathbf{1}}^*\prod_i\chi_{\Rep_i}=\frac{1}{V_G}\sum_g \prod_i\chi_{\Rep_i}\;,
\end{align}
where we have used $(\chi_{\mathbf{1}})^*=1$ with $\mathbf{1}$ the singlet representation in eq.~(\ref{OrTH}). A key assumption so far was distinguishable representations, let us see where the formula above fails when this is not the case. Take for example $G=SO(2)$ and the square of a field $\phi$ in the vector representation,
$$
n_I=\frac{1}{V_{SO(2)}}\int d\mu_{SO(2)}\chi_{\mathbf{1}}^*\chi_\Rep^2=\frac{1}{2\pi}\int_0^{2\pi} d\theta (\chi_\Rep)^2=\int \frac{d\theta}{2\pi} (e^{i\theta}+e^{-i\theta})^2=2\,.
$$
This counting returns two invariants built with $\phi^2$, yet we know there is one only. The offending term is $\epsilon^{ij}\phi_i\phi_j$ which trivially vanishes, so the solution is to take the symmetric product only.

In order to build the character of the symmetric product, it is useful to rotate to the Cartan basis where $g$ is diagonal and $(g_\Rep)_{ii}=x_i$ with $x_i$ parametrising the Cartan subalgebra which generates $[U(1)]^{\rm {rank}\, G}$; for the $SO(2)$ example we would have $x_1=e^{i\theta}$, $x_2=e^{-i\theta}$. Here the action of the group element on the product 
\begin{align}
[g(\phi\otimes\phi')]_{ij}=& (g_\Rep)_{ii}(g_\Rep)_{jj} \phi_i\phi_j'=x_ix_j\phi_{\{i}\phi_{j\}}'+x_ix_j\phi_{[i}\phi_{j]}'\\
\chi_{\textrm{sym}\Rep^2}=&\sum_{i\leq j} x_ix_j
\end{align}
If we had instead three $\Rep$ representations, the character is the ordered sum ($i\leq j\leq k$), and so forth for higher terms. All products can be in fact neatly packed into a single generating function
\begin{align}
    \sum_n\chi_{\textrm{sym}([\Rep]^n)}\phi^n=\prod_i(1+x_i\phi+(x_i\phi)^2+\dots)=\prod_i\frac{1}{1-x_i\phi}=\det[(1-g_\Rep \phi)^{-1}]\,, \label{Phisym}
\end{align}
where we have used $\phi$ as a bookkeeping device, a practice common in the literature and throughout the rest of this work.
This expression can be cast as a Pleythistic exponential, defined as
\begin{align}
\det(1-g_\Rep)^{-1}&=\textrm{Exp}\left(-\textrm{tr}[\log(1-g_\Rep)]\right)=\textrm{Exp}\left(\sum \textrm{tr}(g_\Rep^n)/n\right)\equiv \textrm{PE} [\chi_\Rep](x_i)\,,\\
\textrm{PE}[f](x)&\equiv\textrm{Exp}\left(\sum_{n=1} f(x^n)/n\right)\,.
\end{align}
One can then count singlet representations, i.e. invariants, by the formula
\begin{align}
   \frac{1}{V_G}\sum_g\det(1-\phi g_\Rep)^{-1}= \frac{1}{V_G}\sum_g \PE[\chi \phi](x_i,\phi)=\sum_I n_I \phi^I\;\;.
\end{align}
This expression captures the counting of invariants for any number of representations $\Rep$ marked by the power of $\phi$; i.e. this polynomial in $\phi$ has as coefficient of the $I$'th power the number of invariants with $I$ representations. The name given to this function of $\phi$ is the HS. It gives a name to this operator counting method since it is the central object containing all the information we are after. In order to evaluate the integral for arbitrary $\phi$, it is often useful to use the determinant form and the residue theorem. For our applications however, solving the integral in general is not possible and instead one expands on the field and finds the coefficient of interest.

The extension of the Hilbert series to one more field is straightforward, there will be one more Pleythistic in the integrand
\begin{align}
    H(\phi_\Rep,\varphi_{\Rep'})=n_{k,l}\phi^k\varphi^l=\frac{1}{V_G}\sum_g\textrm{PE}[\phi \chi_\Rep](x_i,\phi) \times\textrm{PE}[\varphi \chi_{\Rep'}](x_i,\varphi) \;,
\end{align}
where $n_{k,l}$ is the number of invariants that can be built off of $k$ $\phi$-fields and $l$ $\varphi$-fields. 

A couple of generalisations will be useful
\begin{itemize}
    \item {\bf Fermions} If the field $\psi$ anti-commutes we should select instead fully antisymmetrised combinations, these are indeed finite and collected in $\det(1+g\psi)$, which when written in terms of an exponential, brings a sign change
    \begin{align}
\textrm{PE}[f]_{\psi}(x)&\equiv\textrm{Exp}\left(-\sum_{n=1} (-1)^nf(x^n)/n\right)\;. \label{PEfer}
    \end{align}
    \item {\bf Flavour} The case in which $n_f$ of our representations are distinct but in the same representation, one has the product of $n_f$ pleythistic exponentials;
    \begin{align}
        \left(\textrm{PE}[f](x)\right)^{n_f}&\equiv\textrm{Exp}\left(n_f\sum_{n=1}  f(x^n)/n\right)\;, \label{PEfl}
    \end{align}
    and one can give the counting as a function of $n_f$.
\end{itemize}
\subsection{Free field case}
The discussion so far did not involve the field nature of our representations, i.e. the spacetime dependence. This field nature is reflected in the fact that $\partial\phi$ is an independent object to build invariants with and one with different transformation properties under the Lorentz group, which for the purposes of counting can be taken as the compactified version $SO(d)$ for $d$ dimensions.

Consider the series of derivatives arranged in an infinite-dimensional array,
\begin{align}
    \OPM_\phi=\left(\begin{array}{cccc}
       \phi\,,& \partial_\mu \phi\,,&  \partial_\mu\partial_\nu\phi\,, &\dots
    \end{array}\right)^T\;.
\end{align}
Let us denote by $\Box$ the vector Lorentz representation (in $SU(2)_L\times SU(2)_R$ notation the $(1/2,1/2)$), the characters of the first two terms in $\OPM$ are $\chi_\Rep$ and $\chi_\Rep\chi_\Box$, yet in the second $\partial_{[\mu}\partial_{\nu ]}=0$. The need to symmetrise a power of representations is something we have encountered before and solved with the PE, so we do here; the sum of traces for any power of derivatives reads
\begin{align}\label{ChiTS}
\chi_{\OPM_\phi}\equiv \sum_n (\partial)^n\chi_{\Rep\times\textrm{sym}(\Box)^n}= \chi_\Rep\sum_n (\partial)^n\chi_{\textrm{sym}\Box^n}=\chi_\Rep\det(1-\partial g_\Box)^{-1} \;,
\end{align}
 where $\partial$ acts here as another bookkeeping device\footnote{In the literature one might find $P=\det(1-\partial g_\Box)^{-1}$ is introduced for succinct notation.}. We have symmetrised on $\partial$, while symmetrising on the representation $\Rep$ is done as in eq.~(\ref{Phisym}), so one might be tempted to (wrongly) assume that to find invariant operators with two fields and an arbitrary number of derivatives we take the group average over
\begin{align}
    (\det(1-\partial g_\Box)^{-1} \phi)^2\mbox{Tr}(g_{\textrm{sym}\Rep^2})\;. \label{wrong1}
\end{align}
This formula, as can be easiest checked in $SO(d)$ with $d=2$ spacetime and a scalar field in $\Rep=\mathbf{1}$, yields 2 for invariant operators of the form $\partial\phi\partial\phi$ and another two as $\phi\partial^2\phi$. The issue encountered is that one not only needs to symmetrise $\partial$ among themselves but also $\partial^n\phi\partial^m\phi$ with $\partial^m\phi\partial^n\phi$. This symmetrisation once more can be achieved with products of the geometric series
\begin{align}\nonumber
    &\prod_i(1+x_i\phi+(x_i\phi)^2+\dots)\prod_j(1+x'_j\partial\phi+(x'_j\partial\phi)^2+\dots)\prod_k(1+x''_k\partial^2\phi+(x''_k\partial^2\phi)^2+\dots)\\
    =&\det(1-g_\Rep\phi)^{-1}\det(1-g_{\Rep\times \Box} \partial \phi)^{-1}\det(1-g_{\Rep\times \textrm{sym}\Box^2}\partial^2\phi)^{-1}\dots=\det(1-\hat g_{\OPM_\phi}\phi)^{-1} \;,
\end{align}
where $x_i$, $x'_i$ are the group element eigenvalues of $g_\Rep$, $g_{\Rep\times\Box}$ etc and the infinite dimensional matrix $\hat g$ is
\begin{align}
g_{\OPM_\phi}\OPM_\phi=&\left(\begin{array}{cccc}
g_{\Rep} & & &\\
& g_{\Rep\times\Box} & &\\
&&g_{\Rep\times\textrm{sym}\Box^2}&\\
&&&\dots
\end{array}\right)\left(\begin{array}{c}
\phi\\
\partial\phi\\
\partial_\mu\partial_\nu \phi\\
\dots
\end{array}\right) & \hat g_{\OPM_\phi}=&\left(\begin{array}{cccc}
1 & & &\\
& \partial & &\\
&&\partial^2&\\
&&&\dots\end{array}
\right)g_{\OPM_\phi} \;,
\end{align}
and the bookkeeping is a bit more involved since every sub-element of $ g_\OPM$ is weighted by a power of $\partial$ as above so eq.~\eqref{ChiTS} is reproduced.
For the  Hilbert series, one can use the determinant form or the PE using the trace-log formula to obtain a function of the character as
\begin{align}
H(q,\phi,\partial)= \frac{1}{V_GV_{SO(d)}}\int d\mu_G\int d\mu_{SO(d)} \textrm{PE}[\phi \chi_{\OPM_\phi}\chi_G](\phi,\partial,x,y)\;,\label{HSraw}
\end{align}
where the right symmetrisation is attained by the PE having as argument $\partial$, not $P=\det(1-g g_\Box)$, i.e. $\PE[\chi](\partial)$ not $\PE[\chi](P)$ which would yield powers of $P$ and lead to eq.~\eqref{wrong1}.
We have obtained an HS that counts all possible operators with any number of fields and derivatives acting on them. This result is already relevant and applicable to QFT while, as we tried to show, it draws solely from the orthogonality theorem with the need to symmetrise giving rise to the use of PE. The extension to fermions and flavour is as sketched in eqs.~(\ref{PEfer}, \ref{PEfl}).

\subsection{IBP, EoM and CFT}
Counting operators in field theory has led naturally to consider the infinite-dimensional representation $\OPM_\phi$ assembled out of field derivatives and hence highly reducible under $SO(d)$. Given it is treated as a whole however the question arises: could an encompassing group have this representation as an irrep? On the other hand, redundancies in the QFT formulation imply that the same physics encoded in the whole set of operators that formula~\eqref{HSraw} counts is described by a subset of them. As math would have it, rather enthrallingly, these two points are connected through the conformal group. 

Let us first pursue the group that would present $\OPM_\phi$ as an irrep. The starting realisation is that the action of the derivative itself is encoded in a raising ladder operator connecting the different Lorentz irreps, take $P_\mu=-i\partial_\mu$
\begin{align}
    i\epsilon_\alpha P^\alpha \OPM_\phi=\epsilon_\alpha \partial^\alpha \OPM_\phi= \left(\begin{array}{cccc}
0 &\epsilon^\mu & &\\
&0 &\eta^{\mu\nu}\epsilon^\rho &\\
&&0&\\
&&&\dots
\end{array}\right)\left(\begin{array}{c}
\phi\\
\partial_\mu\phi\\
\partial_\nu\partial_\rho \phi\\
\dots
\end{array}\right) \;,
\end{align}
and the exponentiated form would shift $\OPM_{\phi(x)}\to \OPM_{\phi(x+\epsilon)}$.
In analogy with spin, we can assume that a ladder operator comes from the complex-coefficient combination of generators
\begin{align}
    P_\mu=& -i\partial_\mu =\frac12\left(L_\mu+i\tilde L_\mu\right) \;,
    & K_\mu=& \frac12\left(L_\mu-i\tilde L_\mu\right)\;,
\end{align}
Now $K$ in matrix form is a lowering ladder operator and as such has the inverse units of $\partial$, that is $[K]=[x]$. Equivalently through its action on spacetime $i\epsilon P$ sends $x\to x+\epsilon$ so $i\varepsilon_\mu K^\mu$ will schematically send $x\to x+x^2\varepsilon$ since $[\varepsilon]=[K^{-1}]=[x^{-1}]$. The explicit expression $K_\mu=c_{\mu\rho\sigma \kappa}x^{\rho}x^{\sigma}\partial^\kappa$ follows; assuming the dimensionless tensor $c$ is made out of the metric one can find its form by demanding $[K,K]$ be closed to be
\begin{align}
    K_\mu=\frac i2\left(x^2\partial_\mu-2x_\mu x\partial\right) \;.
\end{align}
The commutation of $K,P$ does return the action of the Lorentz group ($\mathcal J_{\mu\nu}\equiv ix_{[\nu}\partial_{\mu]}$) but also a new generator
\begin{align}
    [K_\mu,P_\nu]&=i\mathcal J_{\mu\nu}-2i\eta_{\mu\nu}\textrm{D}\;,  & \textrm{D}&\equiv i x^\rho \partial_\rho \;.
\end{align}
The Lie algebra so unearthed is now complete and looking at its entirety one finds themselves contemplating the conformal group $SO(d,2)$. This group has rank one unit higher than $SO(d)$, the extra non-compact element of the Cartan-subalgebra being dilations D with parameter $q$ (in analogy with $x_i$), now crucially a real number rather than a phase.  

The advantages of finding a group in this vast space are twofold, first one recovers character orthogonality for full representations; for this we can split the conformal group Haar measure when compactified ($SO(d+2)$) and in the Cartan basis as
\begin{align}
    &\int d\mu_{SO(d)}\oint_{|q|=1} \frac{dq}{2\pi iq}\prod_{i<r}\left(1-x_iq\right)\left(1-\frac{1}{x_iq}\right)\left(1-\frac{x_i}{q}\right)\left(1-\frac{q}{x_i}\right)\\
    =&\int d\mu_{SO(d)}\int \frac{dq}{2\pi i} \det(1-q g_\Box)[\det(1-q g_\Box)]^* \;,
\end{align}
while the trace in this space includes the parameter $q$ as
\begin{align}
    \chi_{SO(d,2)}=q^{\Delta}\chi_\Rep\det(1-g_\Box q)^{-1}=q^{\Delta}\chi_{\OPM_\phi}(\partial\to q) \;,
\end{align}
replacing the somewhat awkward insertion of $\partial$ in the previous section by a group-defined parameter $q$ which also marks the dimension of our field, $\Delta$. Further, the integral over two characters reads
\begin{align}
    &\frac{1}{V_{SO(d)}}\int d\mu_{SO(d)}\oint \frac{dq}{2\pi iq} \det(1-q g_\Box)[\det(1-q g_\Box)]^*\chi_{SO(d,2)}^*\chi_{SO(d,2)}'\\
    =&\frac{1}{V_{SO(d)}}\int d\mu_{SO(d)} \chi_{\Rep}^*\chi_{\Rep'} \oint \frac{dq}{2\pi i q} q^{\Delta-\Delta'}=\delta_{\Delta\Delta'}\int \frac{d\mu_{SO(d)}}{V_{SO(d)}} \chi_{\Rep}^*\chi_{\Rep'}\;,
\end{align}
so we recover group orthogonality and a parameter that allows the selection of mass dimension in $q$.

It is useful nonetheless to keep track of derivatives, so let us from now on use:
\begin{align}
    \chi_{SO(d,2)}=q^{\Delta}\chi_\Rep\det(1-g_\Box \partial q)^{-1}=q^{\Delta}\chi_{\OPM_\phi}(\partial\to \partial q)\,.\label{tildchifin}
\end{align}

The second advantage of the conformal group, this time a special property rather than a shared feature, is that the derivative operator will be of the ladder form on any representation. As such any combination of fields and derivatives that can be written in the form of a total derivative will feature in the lower components of a $SO(d+2)$ representation but never on the first element, denoted primary operator.  Primary operators are then by construction operators that cannot be given as a total derivative and using $SO(d+2)$ as our group not only counts mass dimension but discards total derivatives implementing integration by parts (IBP) relations.

The orthogonality theorem then can be used to project on singlets of mass dimension $\Delta$ with effective measure (where we substitute $q\to q\partial$ as in eq.~\eqref{tildchifin}
\begin{align}
    &\int \frac{d\mu_{SO(d)}}{V_{SO(d)}}\oint\frac{dq}{2\pi iq} \det(1-q\partial g_\Box)[\det(1-q\partial g_\Box)]^*\chi_{SO(d,2)_{\Delta,\mathbf{1}}}^*\PE[\chi_{SO(d,2)}]\\
    &=\int  \frac{d\mu_{SO(d)}}{V_{SO(d)}}\oint\frac{dq}{2\pi iq} q^{-\Delta} \det(1-q\partial g_\Box) \PE[\chi_{SO(d,2)}]\;. \label{CFTH}
\end{align}
The $q$ integral we will in fact not perform and instead $q$ itself is turned into a bookkeeping device to signal mass dimension as
\begin{align}
H(q,\phi,\partial)= \int \dmu \textrm{PE}[\phi \chi_{SO(d,2)}\chi_G](\phi,q,\partial,\{x_i\})\;,\label{PECFT}
\end{align}
where we have defined the measure
\begin{align}
\dmu\equiv\frac{d\mu_G}{V_G}\frac{d\mu_{SO(d)} }{V_{SO(d)}} \det(1-q \partial g_\Box)\;, \label{measdef}
\end{align}
compared with eq.~\eqref{HSraw} it is the determinant factor addition to the measure that takes care of IBP.

Lastly EoM, or more in general field redefinitions, allow for the removal of certain terms. For a scalar, one can choose terms proportional to $\partial^2\phi$ to be discarded. The conformal group has in fact something to say here too. For representations that saturate the unitarity bound components proportional to $\partial^2\phi$ are to be removed in order to obtain a unitary representation, see~\cite{Henning:2017fpj} for details. Free fields do correspond to unitarity-saturating representations e.g. the scalar
\begin{align}
    \tilde \OPM_\phi=\left(\begin{array}{c}
         \phi  \\
          \partial_\mu \phi\\
          (\partial_{\mu}\partial_\nu -\frac{1}{d}\eta_{\mu\nu}\partial^2)\phi\\
          \dots
    \end{array}\right) \;,
\end{align}
and one can derive
\begin{align}
    \tilde\chi_{\OPM_\phi}=\mbox{Tr}(g_{\tilde\OPM})=\chi_{\OPM_\phi}(1-(q\partial)^2)\,,
\end{align}
since the subtraction of EoM can be thought of as removing $(q\partial)^2$ acting on the representation $\OPM_\phi$.
The objects so obtained are true representations closed under composition so one can decompose a product in terms of $\tilde\chi$ characters, $\tilde\chi_\OPM\tilde\chi_{\OPM'}=\sum_{\rm irreps}\tilde\chi_i$. In this basis however we have lost orthogonality so the formula that counted invariants requires modification. The extra step can be understood with linear algebra and projection on a non-orthogonal basis, considering the equivalence
\begin{align}
    \OPM &\to e_i\;, & e_i\cdot e_j&=\delta_{ij}\;,\\
    \tilde \OPM&\to \tilde e_i\;, & \tilde e_i&=M_j e_j\quad M_{ij}=\tilde e_i\cdot e_j \;,
\end{align}
where Einstein's convention for repeated indices being summed over is used and the magnitude to decompose in the basis is 
\begin{align}
    \PE=& c_i \tilde e_i= c_i M_{ij} e_j \;.
\end{align}
If so the dot product with $e_j$ of $\PE$ returns $c_iM_{ij}$ and one can invert to find 
\begin{align}
    c_l=(M^{-1})_{il}\, e_i\cdot \PE =   e_l\cdot \PE+(M^{-1}-\mathbbm{1})_{il}\, e_i\cdot \PE\equiv H^0+\Delta H \;,
\end{align}
or in a more explicit form
\begin{align}\label{DelHSys}
    \Delta H=\left[\int \frac{d\mu_{SO(d+2)}}{V_{SO(d+2)}} \chi^*\tilde\chi-\mathbbm{1}\right]^{-1}_{ij}\int \frac{d\mu_{SO(d+2)}}{V_{SO(d+2)}} \chi_j \PE[\chi] \;.
\end{align}
One then finds $\Delta H$ from the inversion of the matrix $M$ which itself is determined from the projection of $\chi$ on $\tilde\chi$; given that the characters in both bases only differ for a handful of terms, this matrix can be inverted explicitly to find that $\Delta H$ contains low-dimension (dim $\leq d$) terms.

When the fields to be quantised do not belong in a conformal representation the formula above does not apply and an alternative understanding of the term $\Delta H$ comes into use. In the form language\footnote{Here $d=dx^\mu\partial_\mu$ and we omit $\wedge$ so $dx^i\wedge dx^j\to dx^i dx^j=-dx^jdx^i$.}, IBP would remove operators whose contribution to the action reads
\begin{align}
    \int \mathcal O (dx)^d=\int d\omega_{(d-1)} \;,
\end{align}
which in terms of the Hodge dual $\star  \omega_{d-1}=\tilde \omega_\mu dx^\mu$ reads simply $\mathcal{O}=\partial_\mu \tilde \omega^\mu$. Consider for example $d=4$ and 
\begin{align}\label{eq:om3def}
    \omega_{3\phi}&\equiv \star d\phi=\frac{1}{3!}\varepsilon_{\alpha \mu\nu\rho} \partial^\alpha\phi dx^\mu dx^\nu dx^\rho\;,\\
    d(\omega_{3\phi})&=d(\star d \phi)= \frac{1}{3!}\partial_\beta \partial^\alpha\phi \varepsilon_{\alpha \mu\nu\rho} dx^\beta dx^\mu dx^\nu dx^\rho= -\partial_\alpha (\partial^\alpha \phi)(dx)^d \;.
\end{align} 
From this perspective one can understand the first two terms of $\det(1-g_\Box\partial)=1-\chi_\Box \partial+\mathcal O(\partial^2)$ in eq.~\eqref{CFTH}; the first gives the invariant operators built out of the fields in the PE while the second subtracts those that can be built out of a derivative (with character $\chi_\Box$) times the PE of fields. This second term however picks `valid' forms like $\omega_{3\phi}$ but also vanishing operators, e.g. consider 3 flavours of a scalar field in $d=4$ and 
\begin{align}\omega_3^e=& d\phi_1 d \phi_2 d\phi_3\;, 
&
d \omega_{3}^e=(dx)^4\varepsilon^{\alpha\mu\nu\rho} \partial_\alpha (\partial_\mu \phi_3\partial_\nu \phi_2\partial_\rho \phi_3)=0 \;.&
\end{align} 
This vanishes identically since $[\partial_\mu,\partial_\nu]=0$ yet the product of $\det(1-g_\Box\partial)$ and the PE is not symmetrised in $\partial$ so this term would be picked up by $-\chi_\Box \partial$ in the determinant. While this is the same recurring symmetrisation problem, here it can be tackled differently. In form language this term cancels since $\omega^e_3$ is exact, $\omega^e_3=d \omega_2$, so $d\omega_3^e=d^2 \omega_2=0$. One can correct for this by selecting 2-forms ($d-2$) that would lead to exact 3-forms ($d-1$) and vanishing 4 forms ($d$) by multiplying the PE by $\chi_{\textrm{asym}\Box^2}\partial^2$. This is precisely the third term in $\det(1-g_\Box\partial)=1-\chi_\Box \partial+\chi_{\textrm{asym}\Box^2}\partial^2+\mathcal O(\partial^3)$. There are nonetheless 2 forms which are themselves exact ($\omega_2^e=d\omega_1$) so they give vanishing 3 forms and should be subtracted from the counting of 2-forms. The iteration is so repeated till the spacetime dimension is reached and one reproduces
\begin{align}
    \det(1-g_\Box\partial)= \sum_{n=0}^d(-1)^n \chi_{\textrm{asym}\Box^n}\partial^n\;,
\end{align}
which indeed attains the subtraction of exterior derivatives of non-exact $d-1$ forms of eq.~\eqref{CFTH}. A more in depth study is to be found in ref~\cite{Henning:2015alf} sec 2.2. This formula then returns the correct counting in the absence of EoM, and gives an alternative derivation of eq~(\ref{PECFT}).


When accounting for EoM on the other hand we remove terms that vanish by application of the EoM from the single particle module and hence the PE, PE$[\chi]\to $PE$[\tilde \chi]$ yet the product of the PE and the $\det(1-g_\Box \partial)$ factor (i.e. powers of derivatives) in eq.~(\ref{CFTH}) will give rise to terms that do cancel when EoM are implemented. These are the terms that are not accounted for and require $\Delta H$. Terms in this correction can be profiled by revisiting the problem, $\det(1-g_\Box \partial)$ achieved the subtraction of $d-1$ forms whose exterior derivative does not vanish, and a vanishing exterior derivative was associated with the form being exact $\omega_{d-1}=d\omega_{d-2}$. Now a form can have a vanishing derivative not because of being exact but due to EoM; in form language and for a scalar, not because $d^2\phi=0$ but rather $(dx)^d\partial^2\phi=d\star d\phi=0$. It is these new kinds of forms\footnote{Note that since the PE has been built with the reduced particle module one does not have trivially vanishing forms as $\omega=\omega'\partial^2\phi$} that determine $\Delta H$: closed but non-exact. 

The nature of our example form $\omega_{3\phi}$ as defined in eq.~(\ref{eq:om3def}) changes to closed when using the EoM and is one of the terms present in $\Delta H$, another being $\star F =\varepsilon_{\alpha\beta\rho\sigma} F^{\alpha\beta} dx^\rho dx^\sigma/2$. The presence of these terms in $\Delta H$ and $\det(1-g_{\Box}\partial)$ before and after imposing EoM is sketched in table~\ref{TabDelH}. For $\omega_{2F}$ in particular we use $d\star F\propto \star \partial_\mu F^{\mu\nu} dx_\nu=0$, but we should note that in the same category of constraint, we have $d F=0$. The latter is automatic if $F=dA$ yet in the PE it is $F$, the field strength, that is used to build invariants and no reference to $A=A_\mu dx^\mu$ is explicit.

\begin{table}
\begin{center}
    \begin{tabular}{c|c|c|c|c|c}
         w/o EoM & p-form&&$PE[\chi]$ & $(\det(1-g_\Box \partial)-1)\PE[\chi]$ & $\Delta H$\\  \hline
         $\omega_{3\phi}=\star d\phi
         $ & 3 & $d (\omega_{3\phi})\Rightarrow \partial_\mu \tilde \omega^\mu$  & $\partial^2\phi$ &$-\partial^2\phi$&0 \\
         $\omega_{2F}=\star F
         $ & 2& $d (d\omega_{2F})=0$ & $-\partial^2 F$,& $\partial^2 F$&0\\ \hline
        with EoM & & &$PE[\tilde\chi]$& $(\det(1-g_\Box \partial) -1)\PE[\tilde\chi]$&$\Delta H$\\\hline
        $\omega_{3\phi}=\star d\phi$& 3& $d(\omega_{3\phi})=0$ &0&$-\partial^2\phi$&$\partial^2\phi$\\
         $\omega_{2F}=\star F$&  2&$(d\omega_{2F})=0$&0&$\partial^2F$&$-\partial^2 F$
    \end{tabular}
    \caption{Example with two forms which change to closed when using the EoM and their contribution to $\Delta H$\label{TabDelH}}
\end{center}
\end{table}

 In contrast to the closed form of $\Delta H$ for short CFT irreps,  $\Delta H$ is in general determined by finding the closed but not exact forms. These are postulated to always lead to $\Delta H$ terms of dimension less than that of the spacetime considered~\cite{Henning:2017fpj}; this one can test for themselves by attempting to write higher dimension forms and realising that the closed but non-exact condition implies, loosely speaking, a maximum of one field per form index.   

 The last note to make, now truthfully, is a comment on kinetic terms. After one has postulated the right PE and $\Delta H$ term, the Hilbert series achieves what it has been instructed to do, in particular it removes all terms proportional to $\partial^2\phi$ including $-\partial^2\phi^2=(\partial\phi)^2$, the kinetic term itself. There is then the need for one to add it by hand which is the case both for scalars and fermions. For fields in which the building block we use is not the field itself that we quantise this problem is not present which includes gauge bosons and Nambu-Goldstone bosons. Here for consistency we will always include kinetic terms, this is relevant in particular since in sec.~\ref{sec:modding} we will promote a scalar to an NGB.
 
All in all we can conclude this section by giving the master formula, taking $\Rep$ to encode both internal and Lorentz representation,
\begin{align}
\boxed{H(q,\phi,\partial)= \int \dmu \textrm{PE}[\phi \tilde\chi_{\Rep}\chi_{\Rep}](\phi,q,\partial,x,y)+\Delta \hat H\,.}\label{PEphi}
\end{align}
with $\Delta \hat H=\Delta H+\Delta H_{\rm kin}$ and the case study for this introduction has
\begin{align}
    \textrm{Real singlet scalar:} & &\Delta H=&q^3\partial^2 \phi-q^4\partial^4\,,  &\Delta H_{\rm kin}=&q^4\partial^2\phi^2\,,
\end{align}
while the characters for scalar, spinors and field strengths with $SO(4)\sim SU(2)_L\times SU(2)_R$
\begin{align}
    \chi_{C,{\mathbf 2}}(x_C)=&x_C+x_C^{-1}\;, &  \chi_{C,{\mathbf 3}}(x_C)=&x_C^2+x_C^{-2}+1 \;, \label{chiLor}
    \end{align}
with $C=L,R$ while the Lorentz measure can be restricted to the Cartan sub-algebra
    \begin{align}
    \int \frac{d\mu_{SU(2)_C}}{V_{SU(2)_C}}& =\oint \frac{dx_C}{4\pi i}\left(2-x_C^2-x_C^{-2}\right)\;, \label{HaarLor}
\end{align}
and 
 \begin{align}\label{chiP}
 \det (1-g_\Box \partial)^{-1}&=\left[(1-x_Lx_R\partial)(1-\partial/(x_Lx_R))(1-x_L\partial/x_R)(1-x_R\partial/x_L)\right]^{-1}\;, \\
     \tilde\chi_{0,0}(x_C,q,\partial)&=q\det (1-g_\Box q\partial)^{-1}(1-(q\partial)^2)\;, \\
     \tilde\chi_{1/2,0}(x_C,q,\partial)&=q^{3/2}\det (1-g_\Box q\partial)^{-1}(\chi_{L,\mathbf{2}}(x_L)-(q\partial)\chi_{R,\mathbf{2}}(x_R))\;, \\
     \tilde\chi_{1,0}(x_C,q,\partial)&=q^{2}\det (1-g_\Box q\partial)^{-1}(\chi_{L,\mathbf{3}}(x_L)-(q\partial)\chi_{R,\mathbf{2}}(x_R)\chi_{L,\mathbf{2}}(x_L)+q^2\partial^2)\;, \label{chiF}
 \end{align}
where obtaining the RH representation from the LH $(a,0) \to (0,a)$  is done by exchanging $L\leftrightarrow R$ in the RHS and the chiral field strengths are 
\begin{align}
    F_{L/R}^{\mu\nu}=\frac12\left(F^{\mu\nu}\pm i\varepsilon^{\mu\nu\rho\sigma} F_{\rho\sigma}\right)\,,\label{chiralFS}
\end{align}
which can be used to build the basis in place of $F,\tilde F$.
\subsection{Hidden Symmetry}\label{RevNLHS}
Here we follow the CCWZ procedure \cite{Callan:1969sn, Coleman:1969sm} to later connect it with the linear basis in the next section. Consider a group $G$ with Lie algebra $T=\left\{X_A,t_a\right\}$ with $X$ broken and $t$ unbroken generators and a possible gauged subgroup $\hat G$. The generators $t$ span the group $\mH$ themselves and as such their commutation closes $[t,t]\to t$ so $f_{abA}=0$. For a compact group, the  structure constants can be taken fully anti-symmetric and it follows
\begin{align}
    \left[t_a,t_b\right]&=if_{abc}t_c\;, & \left[t_a,X_A\right]&=if_{aAB}X_B\;.\label{brokLie}
\end{align}
The Nambu-Goldstone fields span the broken space so any transformation $\xi\in G/\mH$ at a space time-point $x$  can be given by a field configuration $\xi(x)$. One could use the exponential parametrisation, 
\begin{align}
\xi&=\textrm{Exp}(i\pi^A(x) X_A),
\end{align}
but other colours are available too.
A relevant difference with linear realisations is that it is not given that $\xi$ is closed under the group action, after all the broken generators that span the group are a subset of a full irrep, the adjoint. To obtain closed transformation properties one can use the factorization of any full group element into a product of broken and unbroken elements to obtain, under a transformation $g$
\begin{align}
    g\, \xi& \equiv \xi'\htr\;, & &\xi'=g\,\xi\,\htr^{-1} \;, \label{xitrans}
\end{align}
where $\htr$ is $\xi$ dependent since for every $\xi$ there is a different unbroken group element required for factorisation. While this achieves a (non-linear) representation, $\htr$'s spacetime dependence implies more structure is needed to construct invariant terms.
Projecting in the Lie algebra one obtains the Maurer-Cartan form
\begin{align}
    \xi^\dagger (\partial_\mu+i \FSL{\mu}) \xi \;, 
\end{align}
with $\FSL{\mu}$ the gauge bosons\footnote{When $F$ denotes a gauge boson we will make the Lorentz index explicit, otherwise is to be understood as a field strength, the abelian case reading $F_{\mu\nu}\sim [\partial_\mu,F_\nu]$.} and $F_\mu=F_\mu^aT_a$ with $T_a$ the generators of $\hat G$. Under a gauge transformation $\hat g(x)$ of $\hat G$, which in general will overlap with broken and unbroken groups, 
\begin{align}
   \xi^\dagger (\partial_\mu+i \FSL\mu) \xi\to \htr\xi^\dagger (\partial_\mu+i \FSL\mu) \xi \htr^{-1}+\htr\partial_\mu \htr^{-1} \;, 
\end{align}
where the usual transformation properties of $\FSL\mu$ under $\hat G$ have been used and $\htr$ now has $x$ dependence both through the transformation infinitesimal parameters and $\xi$. It is useful to separate the Maurer-Cartan into broken $\ngbmc{\mu}=\ngbmc{\mu}^AX_A$ and unbroken $\FSNLub{\mu}=\FSNLub{\mu}^bt_b$
\begin{align}
    \xi^\dagger (\partial_\mu+i \FSL\mu) \xi&\equiv \ngbmc{\mu} +\FSNLub{\mu}\;, \label{MauCa}
\end{align}
since, given the Lie algebra relations of eq.~(\ref{brokLie}), these two transform independently\footnote{
The choice of notation is here geared towards the SM and connecting $a_\mu$ with the photon, {\it not} an axial current.}
\begin{align}
    \ngbmc{\mu}\to & \htr \ngbmc{\mu} \htr^{-1} \;, & \FSNLub{\mu}\to&  \htr \FSNLub{\mu} \htr^{-1}+\htr\partial_\mu \htr^{-1}\;, 
\end{align}
where we note that $\FSNLub{\mu}$ has the transformation properties of a gauge boson and it allows us to define a covariant derivative which renders $D\xi$ covariant under both $G$ and $\htr(\xi)$ action,
\begin{align}
    D_\mu \xi &=(\partial_\mu +i\FSL\mu)\xi-\xi \FSNLub{\mu}\;,  & D_\mu \xi&\to g (D_\mu \xi)\htr^{-1}\;.
\end{align}
In fact it is useful to introduce $\mH_g$, the gauged version of $\mH$, and take $\xi$ is a bi-representation both under $G$ and $\mH_g$. In what we term the CCWZ frame one classifies all fields, including the Maurer-Cartan form $\ngbmc{\mu}$ of eq.~(\ref{MauCa}), by their representation under $\mH_g$ only, hiding $G$.

It follows from the definition that this derivative projects out $\FSNLub{\mu}$ in the Cartan form and
\begin{align}
    \xi^\dagger D_\mu \xi=\ngbmc{\mu}\;.
\end{align}
It is relevant to identify the d.o.f. that this object, a representation under $\mH_g$-only, excites out of the vacuum; to first order in the fields, splitting gauge boson in linear space into unbroken and broken as $\FSL{\mu}\sim\FSLub{\mu}+\FSLbr{\mu}$,
\begin{align}
    \ngbmc{\mu}=iX^A(\partial_\mu\pi_A+\FSLbr{\mu,A})+\mathcal O(\pi^2)\;, 
\end{align}
so simultaneously Goldstones and any gauge bosons that overlap with $G/\mH$, i.e. the broken and massive gauge bosons. This resembles the theory in the unitary gauge but it is not; this description circumvents the need for gauge fixing and the propagator of the massive vector boson differs as it will be elaborated on at the end of sec.~\ref{CCWZnMauCar}.

In general $\ngbmc{}$ is a reducible representation of $\mH_g$ and it can be broken into irreps each treated as an independent field in the PE. As for higher derivatives, while $D$ does not commute with itself, it returns another of the building blocks of the theory, as follows
\begin{align}
    D_\nu (\xi^\dagger D_\mu \xi)&= \partial_\nu \ngbmc{\mu}+[\FSNLub{\nu},\ngbmc{\mu}]=D_\nu \xi^\dagger D_\mu \xi+\xi^\dagger D_\nu D_\mu \xi\\
    &=\frac12\left(\left\{\ngbmc{\nu},\ngbmc{\mu} \right\} +\xi^\dagger D_{\{\nu}D_{\mu\}}\xi+i\xi^\dagger \FSL{\nu\mu}\xi-i\FSNLub{\nu\mu}\right)\;, 
\end{align}
where $\FSL{\mu\nu}$ is the field strength in the linear frame, i.e. $[D_\mu,D_\nu]\Phi_\Rep=i\FSL{\mu\nu}\Phi_\Rep$ with $\Phi_\Rep$ a $G$-only irrrep whereas $a_{\mu\nu}$ is the equivalent for $\mH_g$ irreps $[D_\mu,D_\nu]\phi_r= i\FSNLub{\mu\nu}\phi_r$. The first identity shows that each of the symmetric and antisymmetric combinations project only on the broken algebra, which for the projections on unbroken $\xi^\dagger \FSL{\mu\nu}\xi|_t$ and broken $\xi^\dagger \FSL{\mu\nu}\xi|_X$, $\xi^\dagger \FSL{\mu\nu}\xi=\xi^\dagger \FSL{\mu\nu}\xi|_t+\xi^\dagger \FSL{\mu\nu}\xi|_X$, gives
\begin{align}
    \xi^\dagger \FSL{\mu\nu}\xi|_t&=\FSNLub{\mu\nu}\;, & 
    D_{[\mu}\ngbmc{\nu]}&=i\xi^\dagger \FSL{\mu\nu}\xi|_X\equiv i \FSNLbr{\mu\nu} \;, 
\end{align}
where here and in the above $x_{\{\mu}y_{\nu\}}=x_\mu y_\nu+x_\nu y_\mu$, $x_{[\mu}y_{\nu]}=x_\mu y_\nu-x_\nu y_\mu$.
It is good to emphasize that $D_{[\mu}\ngbmc{\nu]}$ is not zero if there are gauge fields; in their absence it is zero. In practice for our quest to build operators we have that $D_{[\mu}\ngbmc{\nu]}$ is to be traded for another term (a field strength or zero) and in the one particle module it suffices to consider the symmetrised derivative.

For the same purpose, the variation of the kinetic term  will give the key part of the EoM:
\begin{align}
   &\frac{1}{2}\delta\left( \mbox{Tr}\left(\ngbmc{\mu} \ngbmc{}^\mu\right)\right)= \textrm{Tr}\left(\delta( \xi^\dagger D_\mu\xi)\ngbmc{}^\mu\right)=
    -\textrm{Tr}\left(\xi^\dagger\delta\xi|_X D_\mu \ngbmc{}^\mu\right)\;, 
\end{align}
where a few properties like eq.~\eqref{brokLie} have been used.

One has then as a result a one particle module with $D^\mu \ngbmc{\mu}$ terms removed and symmetrised in $D^n \ngbmc{}$'s Lorentz indexes
\begin{align}
    \tilde\OPM_{\ngbmc{}}=\left(\begin{array}{ccc}
        \ngbmc{\mu}\,,&  \frac{D_{\{\nu}\ngbmc{\mu\}}}{2}-\frac{\eta_{\nu\mu}}{d}D^\rho \ngbmc{\rho}\,, &\dots
    \end{array}\right)^T\;, 
\end{align}
the character of this representation is like that of a scalar but without the first term "$\phi$" and corrected for one of the derivatives being $\ngbmc{}$ as
\begin{align}
   \boxed{ \tilde\chi_{NG}(x_C,q,D)=\left(\left[\det(1-g_\Box q D)^{-1}\right](1-q^2D^2)-1\right)\frac{1}{D}\;. } \label{chiNGB}
\end{align}
This is analogous to the short irreps characters of eqs.~(\ref{chiP},\ref{chiF}) and all of them together suffice to span all the particle species in our Hilbert series computation for spin $\leq 1$. In this sense other fields that have appeared already are $\FSNLub{\mu\nu}$ and $\FSNLbr{\mu\nu}$ which are field strengths and so have the conformal characters of $(1,0)$ and $(0,1)$ for $\FSNLub{L,R}=(\FSNLub{}\pm i\tilde{\FSNLub{}})/2$ as in eq.~(\ref{chiralFS})  while they belong in a $\mH_g$-representation. More fields can be added, generically denoted matter, by specifying their spin and $\mH_g$ representation which would also yield their leading interactions with the unbroken gauge bosons in $\FSNLub{\mu}$. 

The counting formula then reads 
\begin{align}
    H=&\int \dmu \textrm{PE}\left[\sum \phi_i \tilde\chi_{\Rep_i}\chi_{\Rep_i}\right]+\Delta \hat H\;, \\
    \sum \phi_i \tilde\chi_{\Rep_i}\chi_{\Rep_i}=&\ngbmc{}\tilde\chi_{NG} \chi_{\Rep^{\mH}_i,\ngbmc{}}+\FSNLub{L}\tilde\chi_{(1,0)}\chi_{\Rep^{\mH}_i,\FSNLub{}}+\FSNLub{R}\tilde\chi_{(0,1)}\chi_{\Rep^{\mH}_i,\FSNLub{}}\\
    &+\FSNLbr{L}\tilde\chi_{(1,0)}\chi_{\Rep^{\mH}_i,\FSNLbr{}}+\FSNLbr{R}\tilde\chi_{(0,1)}\chi_{\Rep^{\mH}_i,\FSNLbr{}}+(\textrm{matter})\;.
\end{align}
In  $\Delta H$ terms made of only matter or field strength can be found from the CFT formula of eq.~(\ref{DelHSys}). For terms with at least one $ \ngbmc{}$ one enumerates closed but not exact forms. These forms will cancel following $D_\mu \ngbmc{}^\mu=0$ with $\ngbmc{}=\ngbmc{\mu} dx^\mu$ and $D_{[\mu}\ngbmc{\nu]}=0$ and are easiest to find in the $\mH_g=\emptyset$ case with $D=\partial$ where the conditions above read $d \ngbmc{}=0$, $d\star \ngbmc{}=0$ and are listed in table~\ref{tab:GBDH}. For the case of non-trivial $\mH_g$ the forms in this table should also be singlets of $\mH_g$ so that $d\omega_p=D\omega_p$.

\begin{table}[h]
    \centering
    \begin{tabular}{c|c|c}
        p-form & p & $\Delta H$ \\ \hline
         $\star\, \ngbmc{i}$ & 3 & $\partial \ngbmc{}$ \\ 
         $\ngbmc{i} \ngbmc{l}  \ngbmc{k}$& 3 & $\partial \ngbmc{}^3$\\  
         $\ngbmc{i} \ngbmc{k}$ & 2 &$-\partial^2 \ngbmc{}^2$\\
         $\ngbmc{i}$&1&$\partial^3\ngbmc{}$\\
         $1$&$0$&$\partial^4$
    \end{tabular}
\qquad\qquad
    \begin{tabular}{c|c|c}
        p-form & p & $\Delta H$ \\ \hline
         $\ngbmc{i} \,\FSNLub{j}$& 3 & $\partial \FSNLub{} \ngbmc{}$\\
         $\ngbmc{i} \star\FSNLub{j}$& 3 & $\partial \FSNLub{} \ngbmc{}$\\
         $\ngbmc{i} \,\FSNLbr{j}$& 3 & $\partial \FSNLbr{} \ngbmc{}$\\
         $\ngbmc{i} \star \FSNLbr{j}$& 3 & $\partial \FSNLbr{} \ngbmc{}$\\
         \end{tabular}
    \caption{Goldstone and field strength closed but non-exact form candidates with $\ngbmc{i}X^i=  \ngbmc{}^\mu dx_\mu$, $\FSNLub{}=\FSNLub{\mu\nu}dx^\mu dx^\nu$. The terms in $\Delta H$ should be $\mH_g$-singlet forms which would narrow them further imposing constraints on the $\mH_g$ indexes $i,j,k$ and promoting $\partial\to D$.}
    \label{tab:GBDH}
\end{table}

One last relevant comment pertains to the fields one uses to quantise and obtain amplitudes. The connection of the field strengths $\FSNLub{\mu\nu}$, $\FSNLbr{\mu\nu}$ to the field strengths in the linear realisation are given by projection and $\xi$ factors, nevertheless, both fields excite the same d.o.f. and are equally valid to obtain the $S$ matrix. Given this equivalence, quantising $\FSNLub{\mu\nu}$ is the algebraically simpler thing to do in the CCWZ Lagrangian. Similarly the subset of $\ngbmc{\mu}$ which contains a broken gauge boson excites a massive vector boson particle so one can again quantise $\ngbmc{\mu}$ directly as a massive spinning particle. This last case has more physical content to it; while in the conventional linear realisation one treats the gauge and NGB separately, in this CCWZ basis one packs them both in $\ngbmc{\mu}$ while its derivatives are split in $\FSNLbr{\mu\nu}$ and $\OPM_{\ngbmc{}}$. This is reminiscent of similar discussions with amplitude methods and points at the CCWZ description as the IR description where different-spin massless particles have unified in a single massive spinning particle. What this CCWZ description obscures and amplitude methods make more evident is however the non-renormalisability of the theory, or in the language of amplitudes the fact that higher spin particles are not elementary, see e.g.~\cite{Arkani-Hamed:2017jhn}.


\subsection{Connection with linear realisation in non-linear coordinates}\label{Connek}

The previous section outlined how to construct the building blocks as representations of $\mH_g$ and count operators in the CCWZ basis. Let us now turn to the formulation of the theory in terms of $G$ representations while in the broken phase. Compared to the CCWZ basis, counting of operators is now not straight-forward in general but elements like the covariant derivative are simpler in structure ($D_\mu=\partial_\mu+i\FSL\mu$) and this setup is the most widespread in the Higgs literature.

Consider a field, in a (in general) reducible linear representation of $G$ with a vev that realises the breaking pattern. The vacuum of the theory is parametrised by
\begin{align}
    U\equiv &\frac{\xi \langle\Phi \rangle}{\sqrt{\langle\Phi^\dagger \rangle\langle\Phi \rangle}}\;,  & \htr&\langle\Phi \rangle=0\label{UDef}\;.
\end{align}
Given $\htr$ annihilating the vacuum, the combination $U$ brings the NGB to the linear representation; this is to say $U$ transforms under $G$ but not under $\mH_g$, $U\to gU$ with no $\htr$ in sight and so its covariant derivative contains no $a_\mu$ and reads
$D_\mu U=(\partial_\mu +iF_\mu)U$.

The field $U$ contains as many d.o.f. as $\xi$ but it is not a complete representation of $G$. This is generically the case for other fields; a linear $G$ representation $\Rep$ is made up of irreps of the broken group $\phi_{r_i}$,
\begin{align}
    \Phi_\Rep=&\sum_{ h \,\rm irreps} \xi \phi_{r_i}\;,  &\left( \begin{array}{cccc}
         \phi_r &\to & \htr\phi_r & \mH_g\\
         \xi & \to & g\xi \htr^{-1} &G\times\mH_g\\
         \Phi_\Rep&\to&g\Phi_\Rep & G
    \end{array}\right)\;, 
\end{align}
However, just like for $U$, $\Phi_\Rep$ might not be a complete representation with terms in the sum missing. Let us consider an illustrative example inspired by the SM at intermediate energies where one can produce bottom quarks but not tops, in particular consider the group breaking and matter content
\begin{align}
\begin{array}{l}
     G=SU(2)\;,   \\
     \mH=U(1) \;, 
\end{array}&& \xi_{(2)}&=\textrm{Exp}\left(i\sum_{A=1,2}\pi^A\sigma_A\right),   & \begin{array}{cc}
     & \mH_g \textrm{\,rep} \;, \\
     \psi_-:& -1/2 \;, \\
     \pi^{\pm}:&\pm 1\;, 
\end{array}&
& \begin{array}{cc}
     & \mH_g \textrm{\,rep} \;, \\
     W^3_\mu:& 0 \;, \\
     W^{\pm}_\mu:&\pm 1\;, 
\end{array}
\end{align}
where the $(2)$ subindex labels the irrep $\xi$ acts on. One can write a linear but incomplete doublet and triplet to bring the fields to a linear representation
\begin{align}
    \Psi&=\xi_{(2)}\left(\begin{array}{c}
         0  \\
         \psi_- 
    \end{array} 
    \right)\;,
    & U_i&=\frac12\textrm{Tr}(\xi_{(2)} \sigma_3 \xi^\dagger_{(2)}\sigma_i)\,,\quad  \vec U=\xi_{(3)}\left(\begin{array}{c}
         0 \\
         0\\ 
         1
    \end{array}\right)\;, \label{CCWZtoLin}
\end{align}
where we will denote by $U$ the $2\times 2$ traceless matrix $U=\sigma_i U_i$.
The $G$-invariant kinetic terms then read
\begin{align}
    -f^2\mbox{Tr}\left(D_\mu U D_\mu U\right)+i\bar\Psi \gamma^\mu D_\mu \Psi \;, 
\end{align}
where $D_\mu \Psi=(\partial_\mu +i\sigma^j W^j_\mu)\Psi$ and $D_\mu U=\partial_\mu U+i[\sigma^j W^j_\mu,U]$ as one has in e.g. the SM. The crucial difference is that one then substitutes $\Psi, U$ with eqs.~\eqref{CCWZtoLin} attaining a linear realisation in non-linear coordinates.

This procedure nonetheless cannot be repeated to count independent operators; take for example the $G$-invariant operator $\epsilon^{ij}\Psi_i \Psi_j$ one can build with $\Psi$, upon substitution of~(\ref{CCWZtoLin}) it vanishes.

For a faithful counting, one can obtain the operators in the CCWZ basis and then rotate to the linear basis with the inverse relation 
\begin{align}
 \phi_r&= \textrm{P}_r\xi^\dagger \Phi_\Rep\;, \label{fromHtoG}
\end{align}
where the projector P$_r$ selects the $\mH_g$ representations present in the spectrum, with
\begin{align}
    \sum_{a} t_a t_a=\sum_iC_i \textrm{P}_i \;, 
\end{align}
with $C_i$ the Casimir of the $i$'th $\mH$ irrep. To convert the derivatives of $\mH_g$ fields one has
\begin{align}
D_\mu\phi_r &=D_\mu P_r \xi^\dagger \Phi_\Rep=P_r\left[(D_\mu \xi^\dagger) \Phi_\Rep+\xi^\dagger D_\mu \Phi_\Rep\right] 
 = P_r\xi^\dagger \left(- \ngbmcL{\mu}+ D_\mu\right)\Phi_\Rep\;, \label{derchan}
\end{align}
with 
\begin{align}
    \ngbmcL{\mu}&\equiv\xi \ngbmc{\mu} \xi^\dagger\;, & \ngbmcL{\mu}\to& G \ngbmcL{\mu}G^\dagger\;.\label{fromutouG}
\end{align}
The combination of $V_\mu$ and $D_\mu$ does remove the charged vector bosons at the linear level
\begin{align}
    -V_\mu+D_\mu=\xi D_\mu\xi^\dagger+D_\mu= \xi\partial_\mu \xi^\dagger-iW_\mu +\xi a_\mu \xi^\dagger+\partial_\mu+iW_\mu=\xi\partial_\mu \xi^\dagger+\partial_\mu +\xi a_\mu \xi^\dagger\,.
\end{align}
The projectors  in the linear basis read
\begin{align}
    \textrm{P}_i^G=\xi P_{i}\xi^\dagger\;, & & &\textrm{P}_i^G\to G \textrm{P}_i^G G^\dagger\;, \label{projg}
\end{align}
and one can trade $D_\mu U$ for $V_\mu$ since both are given in terms of derivatives of $\xi$, in particular one can take
\begin{align}
    V_\mu=&\xi\left(\xi^\dagger D_\mu\xi-\frac{\sigma_3}{2}\mbox{Tr}\left(\xi^\dagger D_\mu\xi\sigma_3 \right)\right)\xi^\dagger\;, \label{Vwoa}
\end{align}
which removes the projection in the unbroken Lie algebra so that $D_\mu$ can be taken to contain $W_\mu$ and not $a_\mu$. 

To illustrate the dangers of using $\Psi$, $U$ as building blocks to count independent operators let us list (fermion-current)$\times$(GB-current) operators in both bases in table~\ref{tab:OpsCCWZvsLin}.
\begin{table}[h]
    \centering
    \begin{tabular}{r|c|c}
    Frame &CCWZ&  Linear\\ \hline
    Fields&$\psi_-, v_{\mu}^\pm$ & $\Psi, U$ \\
    Operators&$\bar\psi_-\gamma^\mu D_\mu \psi_- $& $\bar\Psi i\gamma^\mu D_\mu\Psi\,,\,\bar\Psi \gamma^\mu\Psi \mbox{Tr}(UD_\mu U)\,,\, \bar\Psi \gamma^\mu[U, D_\mu U]\Psi$
    \end{tabular}
    \caption{(Fermion-current)$\times$(GB-current) operators for the CCWZ frame and Linear frame.}
    \label{tab:OpsCCWZvsLin}
\end{table}
Only the fermion kinetic term is allowed for this subsector of operators in CCWZ since $\bar\psi\gamma_\mu \psi v^\mu_\pm$ is not a $U(1)$ invariant. On the Linear frame let us note that $\bar\Psi\gamma^\mu (D_\mu U)\Psi$ vanishes via EOM and IBP. The counting should match on both sides once $\Psi$ and $U$ are subbed in terms of the true d.o.f. For the second term, one can use the property of $U$, $U U=1$
\begin{align}
    \mbox{Tr} ( U D_\mu U)= \frac{1}{2}\mbox{Tr} ( D_\mu (U U) )=0\;, 
\end{align}
the second however requires the substitution in terms of $\xi$
\begin{align}
    UD_\mu U&= \xi\sigma_3\xi^\dagger D_\mu (\xi\sigma_3\xi^\dagger) =\xi D_\mu\xi^\dagger+\xi\sigma_3 \xi^\dagger D_\mu\xi \sigma_3\xi^\dagger=\xi(-v_\mu+\sigma_3 v_\mu \sigma_3)\xi^\dagger\;, \\
    \bar\Psi \gamma^\mu[U, D_\mu U]\Psi&=2\bar\psi_- P_-\xi^\dagger\xi(-v_\mu+\sigma_3 v_\mu \sigma_3)\xi^\dagger \xi P_-\psi_-=2\bar \psi_- P_-(-v_\mu+v_\mu )P_-\psi_-=0\;, 
\end{align}
where $P_-=(1-\sigma_3)/2$. One can then show that the two basis contain the same number of independent operators but after some  non-straightforward manipulation.

From the above discusion one concludes that the straight-forward procedure to obtain the correct counting in the linear frame is,
\begin{itemize}
    \item Compute the Hilbert series in the CCWZ frame.
    \item Rotate the fields to the linear basis with the relations in eq.~(\ref{fromHtoG},\ref{fromutouG}).
    \item In this basis the projectors can be related to the linear representation $U$ in eq.~\eqref{UDef} while instead of $D_\mu U$ it is more convenient to use $V_\mu$ (which inhabits the same $G$ representation and can be obtained as in eq.~(\ref{Vwoa})) and its derivatives.
\end{itemize}

One then would have operators in terms of $\Phi_\Rep\,,\,\ngbmcL{}$, its derivatives in the linear basis and projectors P$_r^G$. 
While this is the procedure that is guaranteed to work, in special cases one can build the truly independent operators fully in the linear basis; a notable example is HEFT whose special properties that allow for this to happen are reviewed in sec.~\ref{LinBasFor}.

\section{Application in HEFT}\label{sec:application-in-HEFT}

The breaking pattern is $G=SU(2)_w\times U(1)_Y\to \mathcal H=U(1)_{\textrm{em}}$ with the whole of the group $G$ being gauged $\hat G= G$.
In this section, we present both the CCWZ and linear formulations as well as a new counting formula in terms of the full group. 

Since it will feature in all cases, let us give the relevant characters
\begin{align}
    \chi_{Y,Q_Y}(x_Y)=&x_Y^{Q_Y}\;,  & \chi_{w,\mathbf{2}}(x_w)=&x_w+x_w^{-1}\;, \\
     \chi_{w,\mathbf{3}}(x_w)=&x_w^2+x_w^{-2}+1\;,  &  \chi_{c,\mathbf{3}}(x_{c,i})=&x_{c,1}+x_{c,1}^{-1}x_{c,2}+x_{c,2}^{-1}\;, \end{align}
     \begin{align}
\chi_{c,\mathbf{8}}(x_{c,i})=&x_{c,1}x_{c,2}+(x_{c,1}x_{c,2})^{-1}+2+x_{c,1}^2x_{c,2}^{-1}+x_{c,2}^2x_{c,1}^{-1}+x_{c,1}x_{c,2}^{-2}+x_{c,2}x_{c,1}^{-2}\;,  & &
\end{align}
with measure
\begin{align} \label{eq:Haarmeasure_internal}
    \int \frac{d\mu_{U(1)_Y}}{V_{U(1)_Y}}&=\oint \frac{dx_Y}{2\pi i x_Y}\;,  \nonumber \\
    \int \frac{d\mu_{SU(3)}}{V_{SU(3)}}&=\frac16\oint \frac{dx_{c,1}dx_{c,2}}{(2\pi i)^2 x_{c,1}x_{c,2}} (1-x_{c,1}x_{c,2})(1-(x_{c,1}x_{c,2})^{-1})\nonumber\\&\qquad\qquad\times(1-x_{c,1}^2x_{c,2}^{-1})(1-x_{c,2}^2x_{c,1}^{-1})(1-x_{c,1}x_{c,2}^{-2})(1-x_{c,2}x_{c,1}^{-2})\;, 
\end{align}
the remaining necessary terms can be obtained, for $U(1)_{em}$ by simple substitution of $U(1)_Y$ expressions, for $SU(2)_w$ borrowing from the Lorentz group in eq.~\eqref{chiLor} while $\chi_{c,\mathbf{\bar 3}}=\chi_{c,\mathbf{3}}(x_{c,1}\leftrightarrow x_{c,2})\;. $
\subsection{The CCWZ frame}\label{sec:frameCCWZ}

The case of HEFT has that $G$ is contained in the gauge group and so gauge fields span all the broken and unbroken Lie algebra with $G=SU(2)_w\times U(1)_Y$ and $\mH=U(1)_{\rm em}$. In the CCWZ basis fields are classified according to their $\mH_g$ representation including the Maurer-Cartan form with Goldstone derivatives. The building of our set of operators is then straightforward but when it comes to quantizing, i.e. choosing the interpolating fields that excite one particle states, we will distinguish two possibilities. The two procedures are physically equivalent and yield the same $S$-matrix and even though one is computationally simpler, we find it illustrative to present both here.

\subsubsection{Quantising Goldstones and Gauge bosons}

The unbroken group direction in the Lie algebra is $t=Q_{\rm em}=T_{3L}+Q_Y$. The generators spanning the broken group can be taken as any other independent linear combination, here we take $X_{1,2}=T_{1,2}$ and $X_z=\cos(\alpha)T_3-\sin(\alpha)Q_Y$, where $\alpha$ parametrises this ambiguity and is unphysical. As a group element $\xi$ can be given in any representation, here let us use a $SU(2)$ fundamental representation with hypercharge $Q_Y$. Using an exponential parametrisation one can write
\begin{align}
    \xi =\textrm{Exp}(-i2 \tan(\alpha) Q_Y\pi_3)\textrm{Exp}(i\vec \pi \cdot \vec\sigma) \;, 
\end{align}
here $T_i=\sigma_i/2$, $\sigma_i$ are the Pauli matrices, and we have rescaled $\pi_3$. One can solve for $\htr(\pi)$ explicitly since the exponentiation can be carried out
\begin{align}\xi=e^{-i2Q_Y\tan(\alpha)\pi_3}\left(\cos({\bpi})+i\vec\pi\cdot\vec\sigma\frac{\sin({\bpi})}{\bpi}\right)\;, \label{xi2y}
\end{align}
where $\bpi=\sqrt{(\pi_i)^2}$ and $\htr(\pi)$ is defined such that, for an infinitesimal action 
\begin{align}
    i(\theta_i T^i +\theta_Y Q_Y)\xi
    -\xi
    i\theta_{\rm em}(\pi)\left(T_3+Q_Y\right)=\delta\pi^j\frac{\partial \xi'}{\partial\pi^j}\;, \label{hdef}
\end{align}
where the unexpanded group element is
\begin{align}
    \htr(\pi)=\textrm{Exp}\left[i\theta_{\rm em}(\pi)(Q_Y+T_3)\right]\;, 
\end{align}
and solving in eq.~(\ref{hdef}) returns
\begin{align}
    \theta_{em}(\pi) =\frac{\theta_Y\tan(\alpha)+ \theta_3\bpi\cot(\bpi)+\pi_3\left(\vec\pi\cdot\vec\theta\right) (1-\bpi \cot(\bpi))/\bpi^2 -\epsilon_{jk3}\theta^j\pi^k}{\tan(\alpha)+\bpi \cot(\bpi)+\pi_3^2(1-\bpi\cot(\bpi))/\bpi^2}\;, \label{thetaem}
\end{align}
which, as it should, reduces to a $\pi$-independent transformation when restricting $g$ to $\mH$. The Maurer-Cartan form decomposition returns, using shorthand notation $s_\alpha=\sin(\alpha)$, $c_\alpha=\cos(\alpha)$,  and tg$_\alpha= \tan(\alpha)$,
\begin{align}
    a_\mu=&\frac{1}{s_\alpha+c_\alpha}\left( c_\alpha \ngbmc{\mu}^Y+s_\alpha \ngbmc{\mu}^3 \right),& \ngbmc{\mu}^z&=\frac{1}{s_\alpha+c_\alpha}(\ngbmc{\mu}^3-\ngbmc{\mu}^Y), & \ngbmc{\mu}^{\pm }&=\frac{1}{\sqrt{2}}\left(\ngbmc{\mu}^1\pm i \ngbmc{\mu}^2\right),
\end{align}
where 
\begin{align}\label{viinpi}
\ngbmc{\mu}^Y&=i(B_\mu-2 \partial_\mu \textrm{tg}_\alpha \pi_3)\;,  & \ngbmc{\mu}^j&=i\,\mathbf{e}_{k}^j\left(2\partial_\mu \pi_k+\mathbf{t}_{k}^l W_{\mu}^l\right)\;, \\
\mathbf{e}&=\delta_{jk}+\left(\frac{s_{\bpi} c_{\bpi}}{\bpi}-1\right) P_{\pi}^{jk}+\frac{s_{\bpi}^2}{\bpi^2}\epsilon_{l jk}\pi^l\;,  &
\mathbf{t}&=\delta_{jk}+\frac{ \bpi c_{\bpi}-s_{\bpi}}{s_{\bpi}} P_{\pi}^{jk}+\epsilon_{l jk}\pi^l\;, \label{uinpi}
\end{align}
with $P_\pi^{jk}=\delta^{jk} -\pi^j\pi^k/\bpi^2$ and we have introduced the vielbien and killing vector, see \cite{Alonso:2016oah,Alonso:2023upf} for details on the geometric treatment of NGB. These components are related as
\begin{align}
    \xi^\dagger D_\mu \xi =Q_Y \ngbmc{\mu}^Y+T_i \ngbmc{\mu}^i=a_\mu Q_{\rm em}+(c_\alpha T_3-s_\alpha Q_Y)\ngbmc{\mu}^z+T_{\pm}\ngbmc{\mu}^\pm\;, 
\end{align}
To sort out the connection between different degrees of freedom it is useful to expand to the linear order to identify which particle excites which field
\begin{align}
   -i (s_\alpha+c_\alpha)\FSNLub{\mu}&= s_\alpha W^3+c_\alpha B+\mathcal{O}(\textrm{field}^2)\;,  \\
    -i(s_\alpha+c_\alpha)\ngbmc{\mu}^z&=  W^3-B+2(1+\textrm{tg}_{\alpha})\partial_\mu \pi_3+\mathcal{O}(\textrm{field}^2) \;, 
\end{align}
so in the Maurer-Cartan form $a_\mu$ there is no Goldstone at the linear level and the combination of gauge bosons appearing in $\ngbmc{z}$ defines the $Z$ boson.
The field strengths that can be obtained both through rotating from linear space $\xi^\dagger F_{\mu\nu}\xi$,
\begin{align}
    \FSNLub{\mu\nu}Q_{\textrm{em}}+w_{\mu\nu}^\pm X_\pm+z_{\mu\nu}X_{z}=&\xi^\dagger\left( Q_YB_{\mu\nu}+T_i W^i_{\mu\nu}  \right)\xi 
\end{align}
but also when restricting to $\mH$ and taking $[D,\ngbmc{}]$, $[D,a]$ as
\begin{align}
i\FSMNLz{\mu\nu}=&\partial_{[\nu}\ngbmc{\mu]}^z\;,  &
    i\FSMNLw{\mu\nu}^\pm=&\partial_{[\mu}\ngbmc{\nu]}^\pm+[a_\mu,\ngbmc{\nu}^{\pm}] \;, & &i\FSNLub{\mu\nu}=\partial_{[\mu} \FSNLub{\nu]}\;, \label{FSinCCWZ} 
\end{align}
which shows these are indeed field strengths admitting a vector field as $[D,\ngbmc{}]$. In explicit form
\begin{align}
    \FSNLub{\mu\nu}=&\frac{1}{c_\alpha+s_\alpha}\left[c_\alpha B_{\mu\nu}+s_\alpha w_{\mu\nu}^3 \right]\;,  &
    \FSMNLz{\mu\nu}=&\frac{1}{c_\alpha+s_\alpha}\left[w_{\mu\nu}^3-B_{\mu\nu}\right]\;, 
\end{align}
with
\begin{align}
    w^i_{\mu\nu}=\left([c_{\bpi}^2-s_{\bpi}^2]\delta_{ij}+2\frac{s_{\bpi}^2\pi_i\pi_j}{\bpi^2}+s_{2\bpi}\frac{\pi^k}{\bpi}\epsilon_{kji}\right)W_{\mu\nu}^j=(\mathbf{e}\cdot\mathbf{t})_{ij}W^j_{\mu\nu}\;.
\end{align}
The kinetic term for massive bosons (except the Higgs) then reads
\begin{align}\nonumber
    \mathcal L_{0}=&-\frac{1}{2g^2}w_{\mu\nu}^+w^{\mu\nu}_--\frac{1}{4}\left(z_{\mu\nu},a_{\mu\nu}\right) n \left(\begin{array}{c}
         z^{\mu\nu}  \\
         a^{\mu\nu} 
    \end{array}\right)-\frac{M_Z^2\det(n)}{2e^2 n_{aa}^2}\ngbmc{z}^2-\frac{M_W^2}{g^2}\ngbmc{+}\ngbmc{-}\\
    &+\frac{\zeta}{4} \ngbmc{\mu\nu}^z\ngbmc{z}^{\mu\nu}+\frac{\zeta'}{2} \ngbmc{\mu\nu}^+\ngbmc{-}^{\mu\nu}+\mathcal L_{gf+gh}\;, \label{BosLag}
\end{align}
where for the last two terms we assume a gauge is chosen that cancels $\zeta,\zeta'$ to linear order and further cancels  $Z\partial \pi$, $W\partial \pi$ couplings. The presence of a $\zeta$ term without gauge fixing will be discussed in the next section. 

The transformation to the mass basis, given
\begin{align}
    \beta_a\equiv&\frac{n_{aa}}{\sqrt{\det(n)}} \;, &\beta_{az}\equiv&\frac{n_{az}}{\sqrt{\det(n)}}\;, 
\end{align}
reads
\begin{align}\label{toMassB1}
    W_{3,\mu}=&(c_\alpha\beta_a-\beta_{az})Z_\mu+A_\mu\;, &\pi_3=&\frac{e \beta_a}{2M_Z(1+\textrm{tg}_\alpha)}\varphi_3\;, \\
    B_\mu=&-(\beta_{az}+s_\alpha \beta_{a})Z_\mu+A_\mu\;, & \pi_\pm=&\frac{g }{2M_W}\varphi_\pm\;,  \label{toMassB2}
\end{align}
so one can use the above to substitute in our building blocks as
\begin{align}
    -i\FSNLub{\mu}
    &= A_\mu-\beta_{az}Z_\mu+\mathcal{O}(\textrm{field}^2)\;, \\ 
-i\ngbmc{\mu}^z
    &=\beta_{a}\left(Z_\mu+e \frac{\partial_\mu\pi_3}{M_Z}\right)+\mathcal{O}(\textrm{field}^2)\;, 
\end{align}
where we note that the appearance of $Z_\mu$ in $a_\mu$ is a generic consequence of the mixing of neutral bosons and in HEFT need not be a small effect, unlike in SMEFT. One cannot have nevertheless that $A$ shows in $\ngbmc{z}$ since it defines the broken gauge boson.
With these substitutions and our choice of gauge fixing the quadratic neutral boson Lagrangian reads
\begin{align}
  \mathcal L_0 = &-\frac{W_{\mu\nu}^+W^{\mu\nu}_-}{2g^2}-\frac{1}{4 e^2}\left(A_{\mu\nu}^2+Z_{\mu\nu}^2\right)
    +\frac12 \partial_\mu\varphi_3 \partial^\mu \varphi_3+\partial_\mu \varphi^+\partial^\mu\varphi^-\\&+\frac{M_Z^2}{2e^2} Z_\mu Z^\mu+\frac{M_W^2}{g^2}W^+W_-+\mathcal L_{gf+gh}'+\mathcal L_{\rm interactions}\;, 
\end{align}
where, with the remaining gauge fixing terms one can have the usual $R_\xi$ gauge. 



Once the rotation to the mass basis has been identified and the propagators are made explicit, the rest of the action will give the interactions and Feynman rules upon substitution of $W,B$ and $\pi$ in the operators built out of Maurer-Cartans and fields listed in tables.~\ref{tab:CCWZmatter}, \ref{tab:CCWZforms} 
in terms of $A,Z,W^\pm$ and $\varphi$ as given in eqs.~(\ref{toMassB1},\ref{toMassB2}).  
The unbroken $U(1)_{\rm em}$ will then dictate the coupling to the photon $A$ but also be accompanied by a host of non-linear terms, for example for the up quark
\begin{align}\nonumber
\bar u\gamma^\mu D_\mu u    =\bar u \gamma^\mu\Bigg(\partial_\mu+i\frac{2}{3}\Big[ &A_\mu-\beta_{az}Z_\mu+\frac{s_\alpha}{c_\alpha+s_\alpha}(\mathbf{e}(\pi)\cdot \mathbf{t}(\pi)-\delta)_{33}(A_\mu+(c_\alpha\beta_a-\beta_{az})Z_\mu)\\
    &+\frac{s_\alpha}{c_\alpha+s_\alpha}(\mathbf{e}(\pi)\cdot \mathbf{t}(\pi)-\delta)_{3j}W^j_\mu+\frac{s_\alpha}{c_\alpha+s_\alpha}(\mathbf{e}(\pi)-\delta)_{3j}2\partial_\mu \pi^j \Big] \Bigg) u\;, 
\end{align}
which yields rather involved Feynman rules and this one just springs from the kinetic term. 

The fact that these non-linear terms do depend on $\alpha$, whereas the quadratic terms and propagators do not, signals some arbitrariness in their form. One could choose a simpler set of scalar coordinates by a field transformation to the equivalent of free-falling coordinates \cite{Cohen:2021ucp} yet even then the computations of amplitudes would be quite involved. As we shall see in the next subsection quantizing different fields that excite the same particle will avoid some of these complications. The merit in the present frame is that the fields do display the effects of a transformation under $G$, so that the broken group is hidden in $\mH$ yet explicitly displayed as in eq.~(\ref{thetaem}), but the fact that our building blocks are $\mH$ representations makes these transformations non-linear and space dependent.  

The counting formula follows from the above discussion and has already been given in ref.~\cite{Graf:2022rco}, 
it reads
\begin{align}
    H=\int \dmu\textrm{PE}\left[\sum \phi_i \tilde\chi_{\Rep_i}\chi_{\Rep_i}\right](\phi_i,x_J,q,D)+\Delta \hat H\;, \label{HilCCWZHEFT}
\end{align}
where the fields we are summing over are listed in \ref{tab:CCWZmatter},\ref{tab:CCWZforms} with the characters for reference are in eqs.~(\ref{chiP}-\ref{chiF},\ref{chiNGB}) and $\Delta \hat H=\Delta H'+\Delta H_{\rm kin}$ given in table~\ref{tab:DHNL}.

\begin{table}[h]
    \centering
    \begin{tabular}{c|c|c|c|c}
        $\mH_g\times SO(1,3)\backslash \phi_i$& $u$ &$d$ &$e$&$\nu$\\ \hline
         $SU(3)_c$& $ \mathbf{3}$  & $ \mathbf{3}$ & $\mathbf{1}$&$\mathbf{1}$\\
         $U(1)_{\textrm{em}}$& $2/3$ &$-1/3$ & $-1$&$0$\\
         $SU(2)_L\times SU(2)_R$ &${}_{(1/2,0)+(0,1/2)}$ &${}_{(1/2,0)+(0,1/2)}$ &${}_{(1/2,0)+(0,1/2)}$ & ${}_{(1/2,0)}$ 
    \end{tabular}
    \caption{Matter in the CCWZ frame}
    \label{tab:CCWZmatter}
\end{table}

\begin{table}[h]
\centering
    \begin{tabular}{c|c|c}
        n-form & n & $\Delta H$ \\ \hline
        $\ngbmc{z} (\,a_{L,R}\,,\,z_{L,R})$& 3 & $\partial \ngbmc{}(a_{L,R} , z_{L,R} )$\\
         $\ngbmc{\mp}\,w^\pm_{L,R}$& 3 & $\partial \ngbmc{\mp}  w^{\pm}_{L,R} $\\
         $\star\, \ngbmc{z}$ & 3 & $\partial \ngbmc{z}$ \\
         $\ngbmc{z}\wedge \ngbmc{+} \wedge \ngbmc{-}$& 3 & $\partial \ngbmc{z} \ngbmc{+} \ngbmc{-}$\\  
         $\ngbmc{+}\wedge \ngbmc{-}$ & 2 &$-\partial^2 \ngbmc{+}\ngbmc{-}$\\
         $\ngbmc{z}$&1&$\partial^3\ngbmc{z}$
    \end{tabular}
    \qquad
\begin{tabular}{c|c|c|c}
    n-form & n & $\Delta H$ &$\Delta H_{\rm kin}$ \\ \hline
     $\star dh$ & 3 &$\partial^2h$ &\\
-&- &-&$\partial^2h^2$\\
         $\psi^\dagger \psi$& 3 & $D\psi^\dagger \psi$&\\
         $\FSNLub{L,R},z_{L,R}$& 2& $-D^2( \FSNLub{L,R},z_{L,R}) $ &\\
          - &- &-&$D\psi^\dagger \psi$
    \end{tabular}
    \caption{Correction term for the HS in the CCWZ frame}
    \label{tab:DHNL}
\end{table}
\subsubsection{Quantising the Maurer-Cartan form and its two form}\label{CCWZnMauCar}

Implicit in our writing of Lagrangian terms above is the assumption that one uses $\pi$, $A$ $W$ as our interpolating fields, i.e. since $\langle 0 |W^{\pm}|$W state$\rangle\neq 0$ we quantise $W\sim a+ b^\dagger$. The invariance of physical observables under field redefinitions however affords us any other field that also satisfies $\langle 0 |\Phi^{\pm}|$W state$\rangle\neq 0$ and it is the case here that selecting others is advantageous, for the $W$ particle at hand this would be $\ngbmc{\pm}$.

For this, we can focus on unbroken group transformations only to find
\begin{align}
    [Q_{\rm em},T_{\pm}]&=\pm iT_{\pm}\;,  & [Q_{\rm em},X_z]&=0\;, 
\end{align}
where one need not specify $X_z$ or in other words all choices of $\alpha$ satisfy the above.
The gauge symmetry that acts on $\mH_g$ irreps need not be specified in terms of NGB to the level of eq.~(\ref{thetaem}), but simply regarded as $e^{iQ_\textrm{em}\theta(x)}$ and the unbroken field strength is
\begin{align}
    i\FSNLub{\mu\nu}=&\partial_{[\mu} \FSNLub{\nu]}\;,  &\langle 0| \FSNLub{\mu}(x)|\textrm{photon}(q)\rangle&=i\varepsilon_{\mu} e^{-iq x}\;, 
\end{align}
while the kinetic term for the $u$ quark reduces to
\begin{align}
    i\bar u \gamma^\mu D_\mu u=i\bar u \gamma^\mu \left(\partial_\mu+\frac{2}{3} \FSNLub{\mu}\right) u\;, 
\end{align}
so one can just as well call $a_\mu$ our photon field. 

As in the previous case nonetheless a term of the form $a_{\mu\nu}z^{\mu\nu}$ would add an admixture of $z$ boson to the electric current, yet now this is entirely dealt with by a linear transformation with no vielbien or metric in sight. Note this does not interfere with gauge invariance  given $\ngbmc{\mu}^z\psi$ transforms covariantly an so does $(\partial_\mu+iQ_{\rm em}\FSNLub{\mu}+\beta_{az}\ngbmc{\mu}^z)\psi$.

Let us focus instead on the differences that considering $\ngbmc{z}^\mu$ the field to quantise rather than $\pi_3$, $Z_\mu$ brings about.  In the above discussion as in SMEFT, one can choose a gauge to cancel out the explicit  $\pi Z$ mixing terms and have $\pi$, $Z$ be the propagating degrees of freedom. The unitary gauge would seem closest to the set-up here since Goldstones are taken to decouple. The present realisation nonetheless does away with gauge fixing and it turns the $\zeta$ term in eq.~\eqref{BosLag} relevant for propagation (neglecting mixing terms)
    \begin{align}
        \mathcal L_z(x)=-\frac1{4(1+\zeta)}(z^{\mu\nu})^2+\frac{M_Z^{2}}{2}\ngbmc{z}^2-\frac{\zeta}{(1+\zeta)}\left(\frac{\partial_{\{\mu} v_{\nu\}}^z}{2}-\frac{\eta^{\mu\nu}\partial v^z}{4}\right)^2 \;, 
    \end{align}
    since both the antisymemtric $z_{\mu\nu}$ and the symmetric $\ngbmc{z}^{\mu\nu}$ have derivatives of the field $\ngbmc{z}$.
The quadratic action in momentum representation
\begin{align}
   \tilde{\mathcal L}_z(q)=-\frac12\ngbmc{z}^\mu\left(\eta_{\mu\nu}\left[q^2- M^2_Z\right]-q^\mu q^\nu\frac{1-\zeta/2}{1+\zeta}\right)\ngbmc{z}^\nu \;. 
\end{align}
The free particle solution has then $q^2=M^2$ and $q_\mu \varepsilon^\mu(\ngbmc{z})=0$, while the propagator reads
\begin{align}
    \Delta_{\mu\nu}(q)=\frac{-i}{q^2-M_Z^2}\left(\eta^{\mu\nu}+\frac{(1-\zeta/2)q^\mu q^\nu}{3\zeta q^2/2-(1+\zeta)M^2_Z}\right) \;, 
\end{align}
and a similar one follows for $W$ bosons. It would be an interesting exercise to compare predictions of this realisation with a modified propagator with those of the quantisation of $Z$, $\varphi_3$ which nevertheless is beyond the scope of this paper.

As for the counting formula, it does not change and reads as in eq.~\eqref{HilCCWZHEFT} where now it is the very building blocks of table~\ref{tab:CCWZforms} that one quantises.

\begin{table}[h]
    \centering
    \begin{tabular}{c|c|c|c|c|c|c|c}
        $\mH_g\times SO(4)\backslash \phi_i$&$h$& $\ngbmc{z}^\mu$ &$\ngbmc{\mu}^{\pm}$ &$\FSMNLz{\mu\nu}$&$\FSNLub{\mu\nu}$ & $\FSMNLw{\mu\nu}^{\pm}$&$G_{\mu\nu}$\\ \hline
         $SU(3)_c$& $ \mathbf{1}$  &$ \mathbf{1}$  & $ \mathbf{1}$ & $\mathbf{1}$&$\mathbf{1}$&$\mathbf{1}$&$\mathbf{8}$\\
         $U(1)_{\textrm{em}}$& $0$ &$0$ &$\pm1$ & $0$&$0$&$\pm 1$&$0$\\
         $SO(4)$ &${}_{(0,0)}$&NG &NG &${}_{(1,0)+(0,1)}$ & ${}_{(1,0)+(0,1)}$&${}_{(1,0)+(0,1)}$ &${}_{(1,0)+(0,1)}$
    \end{tabular}
    \caption{Bosons in the CCWZ frame}
    \label{tab:CCWZforms}
\end{table}

\subsection{Linear frame formula}\label{LinBasFor}

The $G$-only transforming construction $U$ built out of $\xi$ is, using the explicit form in~\eqref{xi2y} acting on a doublet but with hypercharge $Q_Y=1/2$ and selecting $\alpha=\pi/4$ for concreteness,
\begin{align}
    \xi_{(\mathbf{2},1/2)}\left(
    \begin{array}{c}
         0  \\
          1
    \end{array}\right)=e^{-i\pi_3/2}\left(
    \begin{array}{c}
    i\pi^+\frac{\sin\bpi}{\bpi}  \\
         \cos\bpi-i\pi^3\frac{\sin\bpi}{\bpi} 
    \end{array}\right)\equiv U \;. 
\end{align} This transforms with hypercharge $-1/2$ and is a $SU(2)_w$ doublet. It is common in the literature to use the $2\times2$ matrix $\mathbf{U}$ 
\begin{align}
    \mathbf{U}\equiv\left(\epsilon U^*,U\right)=
    \left(\begin{array}{cc}
    e^{i\pi_3}\left(\cos\bpi+i\pi^3\frac{\sin\bpi}{\bpi}\right) &e^{-i\pi_3}i\pi^+\frac{\sin\bpi}{\bpi}  \\
     -ie^{i\pi_3}\pi^-\frac{\sin\bpi}{\bpi}&    e^{-i\pi_3}\left(\cos\bpi-i\pi^3\frac{\sin\bpi}{\bpi}\right) 
    \end{array}\right) \;, 
\end{align}
which transforms as $L\mathbf{U}R$ with $L,R^\dagger$ $SU(2)_w$ and $SU(2)_r$ matrices. This is a useful object when classifying operators in terms of the custodial symmetry. Here we will use instead $U$ and note
and $U\,,U^\dagger$ can be traded for $\mathbf{U}(1\pm\mathbf{T})$ with $\mathbf {T}=\mathbf{U}\sigma_3 \mathbf{U}^\dagger$.

Now the rest of matter fields in the EFT happen to fill up all degrees of freedom needed for linear representations which removes the need for projectors of the form of eq.~(\ref{projg}). The connection between the two bases is explicitly, for a quark doublet and keeping up with the convention of uppercase (lowercase) for fields in the linear (CCWZ) frame,
\begin{align}
    Q_L=\xi_{\Rep_Q} \left(
    \begin{array}{c}
        u_L \\
          d_L
    \end{array}\right)=e^{-i\pi_3/6}\left(\cos\bpi+i\vec\pi\vec\sigma\frac{\sin\bpi}{\bpi}\right)\left(
    \begin{array}{c}
        u_L  \\
          d_L
    \end{array}\right) \;, 
\end{align}
whereas for right handed is as simple as $U_R=e^{i2\pi_3/3}u_R$.

Note that the CCWZ fields can be written, if one introduces the SM Higgs doublet $H$, as e.g. $d_{L}=(H^\dagger Q_L)/|H|$, so that these fields are invariant under the broken part of $SU(2)_w$ which mixes upper and lower components of the doublet and this $d_{L}$ is not the `lower component of the doublet', which is not a rotation invariant definition, but rather one of the two broken-group invariant combinations built out of $q_L$ and $\xi$.

One has then that all matter fields fill $G$ representations but further, the NGB themselves also do even without the $h$ singlet. One can build a full linear representation with the derivative of the GB \footnote{If one generalises the pattern as $SU(N)\times U(1)\to SU(N-1)\times U(1)$ it is only for $N=2$ that the $2N-1$ Goldstones could fill in an adjoint with $N^2-1$ entries} with $\xi$ in the representation $\xi_{(\mathbf{2},0)}$
\begin{align}
   V_\mu=(D_\mu \xi_{(\mathbf{2},0)}) \xi_{(\mathbf{2},0)}^\dagger =[(D_\mu \xi_{(\mathbf{2},0)}) \xi_{(\mathbf{2},0)}^\dagger ]^A T_A \;, 
\end{align}
This does have the same transformation properties and in the same linear space as the construction
 \begin{align}
     \mathbf{V}_\mu= 
     (D_\mu \mathbf{U}) \mathbf{U}^\dagger \;, 
 \end{align}
 which parametrises the triplet space differently in terms of $\pi$ but it is physically equivalent, for this purpose all our expressions in terms of $V$ can be traded for $\mathbf{V}$.
 
 In this basis with $\Psi$, $V_\mu$ $\FSL\mu$ however one has that chiral fermions transform under different groups and no bare mass term is allowed, i.e. $\bar\Psi_R\Psi_L$ is not invariant. This means we need the field $U$ itself. One can treat this field as a spurion in that it would have no derivatives, these already being contained in $V_\mu$. However relations like $U^\dagger U=1 $ mean this does not span the representation it transforms under, and invariants built with $U, U^\dagger$ only are necessarily trivial and field independent. In more familiar terms no potential for the Goldstones is allowed.
 
 The group orthogonality condition allows to project out all invariants built solely out of $U$ with a modified factor in the integrand of the Haar measure that reads
 \begin{align}
     &\left[\int \frac{d\mu_{SU(2)}d\mu_{U(1)}}{V_{SU(2)\times U(1)}\det(1-g U)\det(1-g^\dagger U^\dagger)}\right]^{-1}\frac{1}{ \det(1-g U)\det(1-g^\dagger U^\dagger)} \\&\equiv \frac{N(U,U^\dagger)} {\det(1-g U)\det(1-g^\dagger U^\dagger)}=\frac{(1-U^\dagger U)} {\det(1-g U)\det(1-g^\dagger U^\dagger)} \;.
 \end{align}
The numerator removes all singlets made out of $U$, $U^\dagger$, as follows when integrating this expression with the Haar measure which returns $1$ by design, see \cite{Graf:2022rco} for a discussion of spurions in HS. The product of this factor and the PE however will generate non-trivial invariants and one can define a Hilbert series for $G$ representations as
\begin{align}\label{GloriousFormula}
    H=\int \dmu \frac{1-UU^\dagger} {\det(1-g U)\det(1-g^\dagger U^\dagger)}\textrm{PE}\left[\sum \Phi_i \tilde\chi_i\chi_{G,i}\right]+\Delta\hat H \;, 
\end{align}
where any given term is specified by tables \ref{tab:LinBos},\ref{tab:Linmatter} and $\Delta \hat H=\Delta H + \Delta H_{\rm kin}$ in tables \ref{tab:withU},\ref{tab:DelHG}. As an example taking the LH quark doublet we read from the table its representation under Lorentz and the internal symmetry then build its character as $Q_L\tilde\chi_{(1/2,0)}\chi_{c,\mathbf{3}}\chi_{w,\mathbf{2}}\chi_{Y,1/6}$.
Comparing with (\ref{HilCCWZHEFT}) we see that now we integrate over the whole group $G$  instead of $\mH$ and we introduce non-derivative fields $U,U^\dagger$ without potential.

\begin{table}[h]
    \centering
    \begin{tabular}{c|c|c|c|c|c|c}
        $G\times SO(1,3)\backslash\Phi$&$h$& $V^\mu$ &$U$&$U^\dagger$ &$B_{\mu\nu}$&$W_{\mu\nu}$\\ \hline
         $SU(3)_c$& $ \mathbf{1}$  &$ \mathbf{1}$  & $ \mathbf{1}$& $ \mathbf{1}$ & $\mathbf{1}$&$\mathbf{1}$\\
         $SU(2)_w$&$ \mathbf{1}$  & $\mathbf{3}$  & $ \mathbf{2}$& $ \mathbf{2}$ & $\mathbf{1}$&$\mathbf{3}$\\
         $U(1)_{Y}$& $ 0$  &$0$ &$1/2$ &$-1/2$ & $0$&$0$\\
         $SU(2)_L\times SU(2)_R$ &${}_{(0,0)}$  &NG & Spurion & Spurion$^\dagger$ &${}_{(1,0)+(0,1)}$ & ${}_{(1,0)+(0,1)}$ 
    \end{tabular}
    \caption{Bosons in the linear frame}
    \label{tab:LinBos}
\end{table}

\begin{table}[ht]
    \centering
    \begin{tabular}{c|c|c|c|c|c}
        $G\times SO(1,3)\backslash\Phi$& $Q_L$& $U_R$ & $D_R$  & $L_L$ & $E_R$\\ \hline
         $SU(3)_c$& $ \mathbf{3}$ & $ \mathbf{3}$  & $ \mathbf{3}$ & $ \mathbf{1}$& $ \mathbf{1}$\\
         $SU(2)_w$& $\mathbf{2}$   &  $ \mathbf{1}$ & $ \mathbf{1}$ &$ \mathbf{2}$ &$ \mathbf{1}$\\
         $U(1)_{Y}$& $1/6$ & $2/3$ & $-1/3$  & $-1/2$ & $-1$\\
         $SU(2)_L\times SU(2)_R$ &$(1/2,0)$ & $(0,1/2)$ & $(0,1/2)$ &$(1/2,0)$ &$(0,1/2)$  
    \end{tabular}
    \caption{Matter in the linear frame}
    \label{tab:Linmatter}
\end{table}

\begin{table}
\centering
 \begin{tabular}{c|c|c}
        n-form & n & $\Delta H$ \\ \hline
         $ \star\, U^\dagger V U$ & 3 & $\partial V UU^\dagger$ \\
         $\epsilon^{ijk}V_i\wedge V_l \wedge V_k$& 3 & $\partial V^3$\\  
         $U^\dagger V\wedge V U$ & 2 &$-\partial^2 V^2 UU^\dagger$\\
         $U^\dagger V U$&1&$\partial^3V UU^\dagger$
    \end{tabular}
\qquad\qquad
\begin{tabular}{c|c|c|c}
        n-form & n & $\Delta H$ \\ \hline
         $V W^i,\tilde W^i$ & 3 & $\partial VW$\\
         $V U U^\dagger B $& 3 & $\partial B V UU$\\
         $UU^\dagger V W $& 3 & $\partial UU^\dagger V W$ \\
         $(UU^\dagger)^2 V W$ & 3 & $\partial VW (UU^\dagger)^2$\\
         \end{tabular}
 \caption{HS corrections in the linear frame}
    \label{tab:withU}
\end{table}        

\begin{table}[h]
    \centering
     \begin{tabular}{c|c|c|c}
     n-form & n & $\Delta H$&$\Delta H_{\rm kin} $\\ \hline
      $\star dh$ & 3 &$\partial^2h$ &\\
-&- &-&$\partial^2h^2$\\
         $\Psi^\dagger \Psi$& 3 & $D\Psi^\dagger \Psi$&\\
         $B,WUU^\dagger$& 2& $-D^2( B,WUU^\dagger) $ &\\
         - &- &-&$D\Psi^\dagger \Psi$
    \end{tabular}
    \caption{HS corrections in the linear frame including kinetic terms}
    \label{tab:DelHG}
\end{table}

\subsection{Higgs singlet modding} \label{sec:modding}

The discussion above, both in the CCWZ and the linear basis, included the Higgs boson as indeed the simplest field, spinless and chargeless as in our pedagogical example of sec.~\ref{ReviewHS} eq.~\eqref{PEphi} with $\phi\to h$. 

A class of operators are trivially built with the singlet $h$ by dressing operators without $h$ with a function of $h$; in our Hilbert series, we would have $\mathcal O\to \mathcal O(1+qh+q^2h^2\cdots)$. It is useful to group these in a single operator which is achieved by realising that the $h$-summed series $\mathcal O(1+qh+q^2h^2\cdots)=\mathcal{O}/(1-qh)$ can be collapsed into one as
\begin{align}
    H^h\equiv(1-q h)(H^0+\Delta H)+\Delta H_{\rm kin} \;, \label{modh}
\end{align}
Indeed in theories in which $h$ and $\ngbmc{\mu}$ are all NGB one can find operators such as $\sin(h/f)^2 \ngbmc{z}^2$ where the whole series is known and the term is treated as a single operator. 

Inspired by these NGB instances where $h$ factors in an operator come over a scale $f$ but derivatives are over $\Lambda\sim 4\pi f$ one can further distinguish operators with and without $h$ with a power counting formula. This refinement of the counting in effect reorganizes the series since $(Dh/\Lambda^2)^n$ powers turn into $(D/\Lambda)^n (h/f)^{n}$ and to $\Lambda$-mass dimension $n$ rather than $2n$. At the PE level, we can accommodate the dimensionlessness of $h$ by $h\to h/q$ but further, having modded by $h$ in eq.~(\ref{modh}) one has that only derivatives of $h$ would appear and the substitution $h\to\ngbmc{h}/(q\partial)$ with $\ngbmc{h}=\partial h$ should return a polynomial in $\ngbmc{h}$. This is shown with the determinant form of the PE as:
\begin{align}
 \hspace{-0.5cm}   (1-hq)\PE[h\tilde \chi_{(0,0)}]=&\left(\frac{1-hq}{1-h q }\frac1{\det(1-g_\Box \partial h q^2)}\frac{1-(q\partial)^2 qh}{\det(1-g_{\rm sym\Box} (q\partial)^2 qh )} \times\cdots\right)_{h\to\ngbmc{h}/(q\partial)} \nonumber\\
     = &\frac{1}{\det(1-g_\Box q\ngbmc{h})}\frac{1-q^2\partial\ngbmc{h}}{\det(1-g_{\rm sym\Box^2} (q\partial)\ngbmc{h} )}\times \cdots= \PE[\ngbmc{h}\tilde\chi_{NG}]\;. 
\end{align}
This same equality, using $\det(1-hq)=$Exp$(-\sum(hq)^n/n)$, can be turned into a modified character
\begin{align}\nonumber
    \hspace{-1.5cm} &\left(h\tilde\chi_{(0,0)}(\partial q,g_\Box)-hq\right)_{h\to\ngbmc{h}/\partial q}= \left(hq [\det(1- q_\Box q\partial)^{-1}(1-(\partial q)^2)-1]\right)_{h\to\ngbmc{h}/\partial q}\\
    & =(\det(1- q_\Box q\partial)^{-1}(1-(\partial q)^2)-1)\frac{\ngbmc{h}}{\partial}=\ngbmc{h}\tilde \chi_{NG} \;. 
\end{align}
So in the PE one treats the Higgs as another NGB.  The term $\Delta H$ is modified accordingly, for $h$-only terms
\begin{align}
    \Delta H_{\ngbmc{h}}=& \left[(-q^4\partial^4+q^3\partial^2 h)(1-hq)+\partial^2h^2\right]_{h\to\ngbmc{h}/q\partial}\\
    =&-q^4\partial^4+q^4\partial^3\ngbmc{h}+q^2\partial\ngbmc{h} \;, 
\end{align}
one indeed obtains $\Delta H$  for a single GB as can be checked in tab.~\ref{tab:DelHHiggs} for $\ngbmc{h}\to \ngbmc{z}$-only terms.
There are nonetheless extra terms in $(1-hq)\Delta H$ which are not predicted by the form counting formula, they can be spotted by their absence of $\partial$ and read
\begin{align}
    \Delta H'= -q^4\left(\ngbmc{h} \ngbmc{z} \ngbmc{+}\ngbmc{-}+\ngbmc{z}\ngbmc{h}q^{-2}+5 n_f^2\ngbmc{h}\psi^\dagger\psi+\ngbmc{h}\ngbmc{z}(a_{L,R}+z_{L,R})+\ngbmc{h}\ngbmc{\mp}w^{\pm}_{L,R}\right) \;, 
\end{align}
these terms are not a fluke but represent the difference in the to paths ({\it i}) taking a singlet scalar field, modding for its form factors and changing its mass dimension scaling vs ({\it ii}) starting from its derivative as the building block in the NGB treatment. Indeed through path ({\it i}) the counting formula knows $\ngbmc{h}$ is a total derivative and e.g.
\begin{align}
    \ngbmc{h} \ngbmc{z} \ngbmc{+}\ngbmc{-}=\partial^\mu(\mathcal{F}(h) \ngbmc{z}^\nu \ngbmc{+}^\rho\ngbmc{-}^\sigma \varepsilon_{\mu\nu\rho\sigma})-\mathcal{F}(h)\varepsilon_{\mu\nu\rho\sigma}D^{\mu}\ngbmc{z}^{\nu}\ngbmc{+}^\rho\ngbmc{-}^\sigma \;, 
\end{align}
the antisymmetrised derivatives however return $a_{\mu\nu}$, $w_{\mu\nu}$ and so a different type of operators already counted. Another case is
\begin{align}
   \ngbmc{h}\ngbmc{z}=\partial_\mu(\mathcal{F}(h)\ngbmc{z}^\mu)-\mathcal{F}(h)D_\mu \ngbmc{z}^\mu \;, 
\end{align}
so that now both the EoM and IBP mean this operator can be discarded. Note that in this sense $\ngbmc{z}$ is not in the same footing: $\ngbmc{\mu}^z\neq \partial_\mu \omega_z$, it is a non-abelian GB and cannot be written as a total derivative.

While this formula achieves the right ordering of Higgs insertions, what is taken as the leading order Lagrangian is not as clear-cut a question which has led to much discussion in the literature, see~\cite{Buchalla:2013eza,Gavela:2016bzc}. One can see that the naive application of the counting as above would lead to kinetic terms for GB as dimension 2 but for fermions dimension 4. Here we do not define a leading order term and follow the steps to organize the counting but simply note that the use of spurions would help accommodate any given counting in the Hilbert series.

\begin{table}[h]
    \centering
    \begin{tabular}{c|c|c}
    n-form & n & $\Delta H_{\ngbmc{h}}-\Delta H_{\ngbmc{h}}(\ngbmc{h}=0)$ \\\hline
         $\star\,\ngbmc{h}$ & 3 & $\partial \ngbmc{h}$ \\
         $\ngbmc{h}\wedge \ngbmc{+} \wedge \ngbmc{-}$& 3 & $\partial \ngbmc{h} \ngbmc{+} \ngbmc{-}$\\  
         $\ngbmc{h}\wedge \ngbmc{z}$ & 2 &$-\partial^2 \ngbmc{h}\ngbmc{z}$\\
         $\ngbmc{h}$&1&$\partial^3\ngbmc{h}$\\
         $\ngbmc{h}a_{L,R},z_{L,R}$ &3 & $\partial\ngbmc{h}a_{L,R},z_{L,R}$
    \end{tabular}
    \caption{In order to obtain $\Delta H_{\ngbmc{h}}$ one takes $\Delta H$ and replaces all terms with the Higgs $h$ by the terms above.}
    \label{tab:DelHHiggs}
\end{table}

Finally let us summarise the different counting formulas that follow from the different assumptions

\begin{itemize}
    \item Straight-forward mass dimension Hilbert series
    \begin{align} \label{eq:HS_mass_dim}
        H=& \int\dmu  \PE\left[\sum\phi_i\tilde{\chi}_{\Rep_i}\chi_{\Rep_i}\right] +\Delta \hat H
        = H_0+\Delta H+\Delta H_{\rm kin} \;. 
    \end{align}
    \item Higgs form-factor-resummed or modded form
    \begin{align}\label{eq:heft_modded}
        H^h
        &=(1-hq)(H_0+\Delta H)+\Delta H_{\rm kin} \;.
    \end{align}
    \item Higgs as GB or chiral counting
    \begin{align} \label{eq:HS_chpt}
        H^{\overline{\ngbmc{h}}}=& \left((1-hq)(H_0+\Delta H)+\Delta H_{\rm kin}^h\right)_{h\to\ngbmc{h}/\partial q}+\Delta H_{\rm kin}^\psi \\
        =&
        \int \dmu \textrm{PE}[\ngbmc{h}\tilde{\chi}_{NG}+\sum_{\phi\neq h} \phi_i\tilde{\chi}_{\Rep_i}\chi_{\Rep_i}] +\Delta H_{\ngbmc{h}}+\Delta H'+\Delta H_{\rm kin}^\psi \;. 
    \end{align}
\end{itemize}

\section{Technical details on the code}\label{sec:code}

\begin{figure}[h]
    \centering
    \includegraphics[scale=0.8]{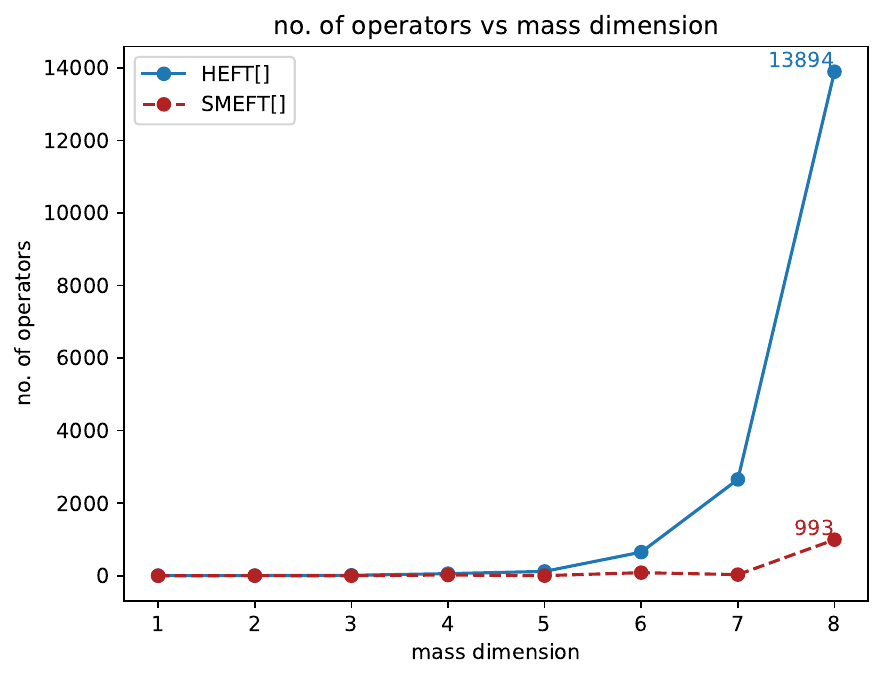}
    \caption{In this plot we have shown how the number of operators changes for a single flavor $(n_f=1)$ by changing the mass dimension}
    \label{fig:counting}
\end{figure}

In this section, we discuss automating the procedure mentioned earlier and briefly describe the available Mathematica code \href{https://github.com/shakeel-hep/HEFT_HS}{in this link}. The primary goal of the code is to compute the Hilbert series for HEFT. To achieve this, we have standardised the notation for all field variables. A list of field names will be printed upon loading the main program, \texttt{code.m}. The code requires two primary inputs: the representation of fields under the gauge and Lorentz groups, which we have provided in Section~\ref{sec:application-in-HEFT} and the Haar measure of these groups shown in eq.~\eqref{eq:Haarmeasure_internal}. To increase the generality of this code, we aim to merge it with \texttt{GrIP} \cite{Banerjee:2020bym} in the near future. Upon successful code loading, the following commands can be executed: \texttt{SMEFT[dim]} and \texttt{HEFT[dim]}. These commands generate HS at the mentioned mass dimension (\texttt{dim}) where the power counting is done in terms of the canonical mass dimension of the operators, as defined by eq.~\eqref{eq:HS_mass_dim}. However, phenomenological studies do not typically use the canonical mass dimension counting for HEFT. A function of the physical Higgs field, $\mathcal{F}(h) = 1 + h + h^2 + h^3 +\cdots$, is usually factored out from the HEFT operators. This obscures the mass dimension counting and introduces redundancy. Section~\ref{sec:modding} discusses this issue and its systematic implementation. To generate Hilbert series output in this setup, use the command \texttt{HEFTModded[dim]}. The code implements this according to eq.~\eqref{eq:heft_modded}. Another phenomenologically motivated operator basis construction requires the physical Higgs field to be treated as a pseudo Nambu-Goldstone boson, the HS output in this arrangement can be followed from eq.~\eqref{eq:HS_chpt}. The \texttt{HEFTngb[dim]} command assists in this scenario. The \texttt{HEFTLinear[dim]} command facilitates the calculation of HS output within the full symmetry group by employing the spurions $U,U^\dagger$. Additionally, we have a function that aids in the counting of the operators named \texttt{Counting[expr]}, it provides the number of total operators as well as an operator class-wise counting. Table~\ref{tab:commands} summarizes the commands and their functionalities.
\begin{table}[t]
\centering
\renewcommand*{\arraystretch}{1}
\begin{tabular}{c|c|l}
        Commands & Input & Description \\ \hline
         \multirow{4}{*}{\texttt{SMEFT[n]} }& \multirow{16}{*}{\texttt{n }$\in \mathbb{N}$} & Computes the Hilbert series for SMEFT for a given \\
         & & mass dimension dimension = \texttt{n}. The power counting   \\ 
         & & scheme is determined by the mass dimension of the \\ 
         & & operators.\\
         & & \\
         \multirow{3}{*}{\texttt{HEFT[n]} }&  & This generates HEFT output using a power counting \\
         & &analogous to SMEFT, based on the mass dimension  \\
         & & of operators. \\
         & & \\
         \multirow{2}{*}{\texttt{HEFTModded[n]} } &  & Generates HEFT operators with all non-derivative \\
         & & insertions of the Higgs $h$ counted as one. \\
         & & \\
         \multirow{2}{*}{\texttt{HEFTngb[n]} }&  & Generates HEFT operators with the $\mathcal{F}(h)$ modded and \\
         & &treats the physical Higgs field $(h)$ as a Goldstone.\\
         & & \\
         \multirow{2}{*}{\texttt{HEFTLinear[n]} }&  & Generates output in unbroken group with \\
         & &  explicit presence of the spurion $U(U^\dagger)$ \\
         \hline
         \multirow{2}{*}{\texttt{Counting[expr]} }& \multirow{2}{*}{\texttt{HS output}} & Counts number of operators in an HS output, \\
         &  & provides class-wise counting as well as total counting.\\
         \end{tabular}
 \caption{Summary of the useful commands. Here, $\mathbb{N}$ denotes the set of natural numbers.}
    \label{tab:commands}
\end{table}
Next, we present an analysis of the operator counting at each mass dimension generated by our code. To cross-reference our findings with the results in \cite{Graf:2022rco}, we have implemented a replacement rule, \texttt{XcheckList}, which maps our operator basis to the form presented in eq.~(5.16) of \cite{Graf:2022rco}. A key discrepancy arises in counting operators involving $L \bar{L}$ terms. This difference can be attributed to the explicit inclusion of a right-handed neutrino in the operator counting of \cite{Graf:2022rco}. At dimension 4, we also incorporate the kinetic terms of the fields, resulting in a few extra operators. All differences can be accounted for and the two countings equated if in a given formula the RH neutrinos can be set to zero (i.e. have not been grouped together under a lepton label), and indeed we have checked that our results agree with those of \cite{Graf:2022rco}. In eq.~\eqref{eq:count} a summary of the operator counting, including the flavor variable $n_f$, is provided. This counting is performed with mass dimension based power counting using the command \texttt{HEFT[dim]}.
\begin{align}\label{eq:count}
    dim1: &\; 1\;, \nonumber\\
    dim2: &\; 3\;, \nonumber\\
    dim3: &\; 3+n_f+7 n_f^2\;, \nonumber\\ 
    dim4: &\; 27+n_f+29\; n_f^2\;, \nonumber\\
    dim5: &\; 30+2\; n_f+87\; n_f^2\;, \nonumber\\
    dim6: &\; 181+2\;n_f+(1429\; n_f^2)/4-(3\; n_f^3)/2+(445\; n_f^4)/4\;, \nonumber\\
    dim7: &\; 335 + (6669\; n_f^2)/4 + (3\;n_f^3)/2 + (2601\;n_f^4)/4\;, \nonumber\\
    dim8: &\; 2336 - (21\; n_f)/2 + (28185\; n_f^2)/4 + (27\; n_f^3)/2 + (18035\; n_f^4)/4 \;\; .
\end{align}
\noindent
To expand on the operator basis presented in eq.~(5.16) of Ref.~\cite{Graf:2022rco}, we use our code to generate the HS for the HEFT dimension-6 case, as shown in eq.~\eqref{eq:heft6}. We have also adopted the replacement rules defined in eq.~(5.14) of the same reference to enhance brevity. Those rules are implemented within our code through \texttt{XcheckList}.
\begin{align}\label{eq:heft6}
H_6 = &\;\; h^6+2 h^4 V^2+ 2n_f^2\; h^3 L \Lbar + 68 V^2 X^2+6 X^3 + 5 h^2 V^4+4 h^2 V^3\mathbb{D} +4 h^2 V^2\mathbb{D}^2 + \nonumber \\
& 4 n_f^2\; h^3  Q \Qbar+n_f^2\;  h^2 (L^2+\Lbar^2)  V+5 n_f^2\; h^2 L \Lbar V+ 8 n_f^2\;h^2  Q \Qbar V+ 72 n_f^2 Q \Qbar V X+\nonumber \\ 
& 6n_f^2\; (L^2+\Lbar^2) V^3+3 n_f^2\; (L^2 + \Lbar^2)  V^2\mathbb{D}+8 n_f^2\;(L^2 + \Lbar^2)  V X+ 6n_f^2\; h L \Lbar  X+   \nonumber \\ 
& 34n_f^4\; L \Lbar Q \Qbar+23n_f^2\; L \Lbar V^3+13n_f^2\;  L \Lbar V^2 \mathbb{D}+ 36 n_f^2 \; L \Lbar V X+\left(8 n_f^4+n_f^3\right)(L^2 + \Lbar^2) Q \Qbar+\nonumber \\ 
& 3n_f^2\; h (L^2 + \Lbar^2) V \mathbb{D} +\left(2 n_f^2-n_f\right)\; h (L^2+ \Lbar^2)  X+10 n_f^2\; h L \Lbar V^2+8n_f^2\;  h L \Lbar V\mathbb{D} +\nonumber \\ 
&  20 n_f^2\; h Q \Qbar V^2+16 n_f^2\; h Q \Qbar V\mathbb{D}+16 h n_f^2 Q \Qbar X+36 n_f^2 Q \Qbar V^3+20 n_f^2\; Q \Qbar V^2\mathbb{D}+\nonumber \\ 
& \left(30 n_f^4+6 n_f^2\right) Q^2 \Qbar^2+10 V^6+10 V^5 \mathbb{D}+5 V^4\mathbb{D}^2 +32 V^4 X+16  V^3 X \mathbb{D} \nonumber \\ 
& 8 h^2 V^2 X + 10 h^2 X^2+\left(\frac{9 n_f^2}{2}+\frac{3 n_f}{2}\right) h (L^2 + \Lbar^2) V^2+ \left(\frac{n_f^2}{2}+\frac{n_f}{2}\right) h^3 (L^2 + \Lbar^2)+\nonumber \\ 
& \left(\frac{n_f^4}{12}-\frac{n_f^2}{12}\right) (L^4 + \Lbar^4) + \left(\frac{3 n_f^4}{2}+ \frac{n_f^3}{2}\right) (L^3 \Lbar + L \Lbar^3)+  \left(\frac{19 n_f^4}{4}+\frac{3 n_f^3}{2}+\frac{7 n_f^2}{4}\right) L^2 \Lbar^2 +\nonumber \\ 
& \left(\frac{15 n_f^4}{2}-\frac{3 n_f^3}{2}\right) (LQ^3 +\Lbar\Qbar^3)+ \left(\frac{25 n_f^4}{6}-\frac{3 n_f^3}{2}-\frac{2 n_f^2}{3}\right) (L\Qbar^3+\Lbar Q^3)\;\;. 
%
\end{align}
\noindent
We plot in Fig.~\ref{fig:counting}, how the number of independent operators grows with increasing mass dimension for different EFTs and power counting.

\section{Summary}
Characterising the scalar boson discovered at the LHC is the key to complete our understanding of electroweak symmetry breaking. Current and future experimental data requires for its interpretation a theory as general as possible so as to leave no possibility unexplored. EFT offers this generality when one counts all operators to a given order. This counting problem has a known solution in terms of the Hilbert Series, including the case of spontaneous symmetry breaking necessitated for application to EWSB. The theory framework to describe SSB, `phenomenological' Lagrangians a.k.a. CCWZ, has long been known yet in the most general EFT to describe EWSB Lagrangians in a different frame, the linear one, has been traditionally used. In this sense the use of the CCWZ frame became evident when an unambiguous counting formula was first made explicit in~\cite{Graf:2022rco}.

The present work explored the counting and connection between the CCWZ and linear frames in general and for HEFT in particular. Either frame captures the same physics so the researcher can choose at their own discretion yet we found each has certain advantages. The CCWZ frame is the frame for unambiguously counting operators and presents fields which pick up a phase in the unbroken group under a broken rotation but do not mix with one another. The linear frame displays the limit of a renormalisable theory most clearly yet the perturbative expansion obscures the broken-gauge invariance of physical states.  For the case of HEFT we found it possible to perform the counting in the linear frame, presented the new formula in eq.~\eqref{GloriousFormula} and underlined that it is the special property of filled up linear representations (including the Maurer-Cartan form in the linear frame) that affords this. Lastly, this work presented a Mathematica code to count operators in HEFT for the three counting formulae of eqs.~(\ref{eq:HS_mass_dim}, \ref{eq:heft_modded}, and \ref{eq:HS_chpt})  as well as SMEFT to any order provided enough computer time.

\appendix 
\acknowledgments
This work is carried out under the support of STFC under grants ST/V003941/1 and ST/X003167/1.










\bibliographystyle{jhep}
\bibliography{ref.bib}

\end{document}